\documentclass{aa}   
\usepackage{graphicx}
\usepackage{natbib}
\bibpunct{(}{)}{;}{a}{}{,}    
\usepackage{txfonts}

\begin{document}
   \title{Stability and Formation of the Resonant System HD~73526}
    
   \author{Zs. S\'andor, 
          \inst{1,2} 
          W. Kley,\inst{2} 
	  \and 
	  P. Klagyivik\inst{1} 
          } 
 
   \offprints{Zs. S\'andor} 
 
   \institute{Department of Astronomy, E\"otv\"os Lor\'and University, 
              P\'azm\'any P\'eter s\'et\'any 1/A, H--1117 Budapest, Hungary\\ 
              \email{Zs.Sandor@astro.elte.hu} \email{P.Klagyivik@astro.elte.hu} 
         \and 
            Institut f\"ur Astronomie und Astrophysik, Abt. Computational Physics,  
	    Universit\"at T\"ubingen,\\ 
            Auf der Morgenstelle 10, D--72076 T\"ubingen, Germany\\ 
	     \email{kley@tat.physik.uni-tuebingen.de} 
	     }

\date{Received ; accepted } 
 
\abstract 
   { 
   Based on radial velocity measurements it has been found recently that the two 
   giant planets detected around the star HD~73526 are in 2:1 resonance. However,  
   as our numerical integration shows, the derived orbital data for this system result  
   in chaotic behavior of the giant planets, which is uncommon among the resonant 
   extrasolar planetary systems. 
   } 
   { 
   We intend to present regular (non-chaotic) orbital solutions for the giant planets  
   in the system HD~73526 and offer formation scenarios based on combining 
   planetary migration and sudden perturbative effects such as planet-planet scattering  
   or rapid dispersal of the protoplanetary disk. A comparison with the already  
   studied resonant system HD~128311, exhibiting similar behavior, is also done. 
   } 
   { 
   The new sets of orbital solutions have been derived by the Systemic Console  
   (\texttt{http://www.oklo.org}). The stability of these solutions has been investigated 
   by the Relative Lyapunov indicator, while the migration and scattering effects are 
   studied by gravitational N-body simulations applying non-conservative forces as 
   well. Additionally, hydrodynamic simulations of embedded planets in 
  protoplanetary disks are performed to follow the capture into resonance. 
   }  
   { 
   For the system HD~73526 we demonstrate that the observational 
  radial velocity data are consistent with a coplanar planetary 
  system engaged in a stable 2:1 resonance exhibiting apsidal corotation. 
   We have shown that, similarly to the system HD~128311, the present dynamical  
   state of HD~73526 could be the result of a mixed evolutionary process melting 
   together planetary migration and a perturbative event. 
   } 
   {} 
 
 
 
\maketitle 
 
%
 

\section{Introduction} 
 
Nearly one third of the multiplanet extrasolar planetary systems contain pairs of 
giant planets engulfed in mean motion resonances. Resonant extrasolar planetary 
systems have special importance when studying the formation of extrasolar planetary 
systems as they require a dissipative mechanism operating on the 
planets which is capable of changing the semi-major axis of their orbits.   
Hence, resonant planetary systems may support the planetary migration scenario, 
and thus help in answering the question why Jupiter-sized giant planets do not orbit 
at their formation place. 
According to the classical formation theories such as planetesimal accretion followed 
by core instability or the gravitational instability model for the giant planets, 
it is generally accepted that they formed quite far from their host star  
(beyond the snowline, which is $a\sim 3-4$ AU for a Sun-like star), 
where volatile elements can condense and accrete 
\citep[e.g.][]{Hayashi1981PThPS,Pollack1984ARA&A,Perryman2000RPPh}. 
On the other hand, observations show that a large fraction of Jupiter-sized planets 
are found very close to their host star at distances smaller than $a=1$AU.  
 
Planetary migration occurs when the planets are still embedded into a protoplanetary 
disk and, due to tidal interactions between the planets and the disk, their semi-major  
axis decrease. During the migration process, a resonant capture can occur between 
two planets, if certain dynamical conditions are fulfilled. Once the capture happens 
and the orbital decay is sufficiently slow (adiabatic), the resonant configuration is 
preserved during the migration, and the two planets can travel very close to their host 
star. The very efficiency of this mechanism is clearly demonstrated by the well 
known resonant system around the star GJ 876. The giant planets in this system are 
engaged in a 2:1 resonance, having very well determined elements due to their short 
orbital periods ($P_1\approx 30$, $P_2\approx 60$ days). It is true in general cases 
\citep{Beaugeetal2006MNRAS} and also in the particular case of GJ 876  
\citep{Lee&Peale2002ApJ,Kleyetal2005A&A}, that as a result of an adiabatic 
convergent migration process, beside the mean motion resonance the orbits of the 
giant planets also exhibit apsidal corotation, (or in other words apsidal resonance), 
where the osculating orbital ellipses of the giant planets rotate with the same mean 
angular velocity.    
 
Numerical integrations based on recently observed orbital data of the planetary 
system HD~128311 \citep{Vogtetal2005ApJ} show that the giant planets are in a 2:1 
mean motion resonance without exhibiting apsidal corotation. In a recent study, 
\citet{Sandor&Kley2006A&A} offered a mixed evolutionary scenario for this system 
combining an adiabatic migration leading the system into a 2:1 resonance and a 
sudden perturbation. This perturbation could be either a close encounter between one 
of the giants and a relatively small mass ($\sim 10 M_{\oplus}$) planet already 
existing in the system (which will refer to in the following as planet-planet scattering), 
or the fast termination of the planetary migration due to a sudden dispersal of the 
protoplanetary disk. Both these perturbations can be strong enough to induce 
relatively large-amplitude oscillations of the eccentricities and the resonant angles.  
While retaining the mean motion resonance, the planet-planet scattering may even 
break the apsidal corotation between the orbits of the giant planets.  
 
In this paper we will support the above described mixed evolutionary scenario by  
the detailed analysis of the newly discovered resonant system HD~73526. 
In a recent paper \citet{Tinneyetal2006ApJ} reported that the giant planets of 
HD~73526 are in the 2:1 resonance. It has been found that only the resonant angle 
$\Theta_1 = 2\lambda_2 - \lambda_1 - \varpi_1$ librates, while both $\Theta_2 = 
2\lambda_2 - \lambda_1 - \varpi_2 $ and $\Delta \varpi = \varpi_2 - \varpi_1$ 
circulate. 
Here, $\lambda_{1,2}$ denote the mean longitude of the inner and outer planet, and 
$\varpi_{1,2}$ their periastron longitude. However, according to our numerical 
integration, the recently published orbital data by \citet{Tinneyetal2006ApJ}, 
as shown in Table~\ref{table:1}, result in 
a weakly chaotic and irregular behavior of the system. 
 
Since chaotic behavior is uncommon among the known resonant extrasolar planetary 
systems, and may not guarantee the stability of the giant planets for the whole lifetime 
of the system, we have searched for regular orbital solutions for the giant planets as 
well. 
By using the Systemic Console (\texttt{http://www.oklo.org}), we have found new 
{\it coplanar} orbital solutions for the giant planets around HD~73526 exhibiting 
regular behavior. The regular nature (and thus the long-term stability) of these orbits 
have been checked carefully by numerical integrations and the Relative Lyapunov 
indicator chaos detection method \citep{Sandoretal2004CeMDA}. Numerical 
integrations based on our newly derived orbital data exhibit very similar behavior of 
the eccentricities of the planets in HD~73526 to those around HD~128311.  
 
This paper aims at a complete analysis of the resonant system HD~73526 including 
the stability investigation of the new orbital solutions and modeling the formation of 
the resonant system HD~73526. The paper is structured as follows: first we present 
four sets of new orbital data. Then we investigate the stability of these orbital 
solutions by using the Relative Lyapunov indicator chaos detection method, and 
finally, we offer possible evolutionary scenarios for the system, which are similar to 
the studied ones in the case of HD~128311.

\section{Original orbital data and their stability} 
 
It has been mentioned also in the case of HD~128311 by \citet{Vogtetal2005ApJ}, 
that a kinematic orbit fit for orbital parameters in the case of a strongly interacting 
resonant pair of giant planets based only on two unperturbed ellipses, often results in 
a destruction of the system within a few thousand years. This is true for HD~73526 
as well. 
 
Thus \citet{Tinneyetal2006ApJ} have also carried out a self consistent three-body 
dynamical orbit fit to the observed velocities (as listed in Table 1). They found this 
dynamical fit to be stable over 1 Myr integration time. Based on these results of 
\citet{Tinneyetal2006ApJ}, we have also performed numerical integration of the 
system. We have found that the giant planets around HD~73526 are indeed in a 2:1 
mean motion resonance having libration in the resonance variable $\Theta_1$ with 
an amplitude $\sim 90^{\circ}$. 
 
However, the behavior of the eccentricities is irregular, see Figure 1. This means  
that the system is chaotic, which of course may not exclude a practical stability of the 
giant planets. 
However, based on the behavior of the observed resonant systems to date, 
we would prefer orbital solutions exhibiting regular orbits showing  
a clear dynamical stability for the whole life-time of the system.  
 
\begin{figure} 
   \centering 
   \includegraphics[width=8cm]{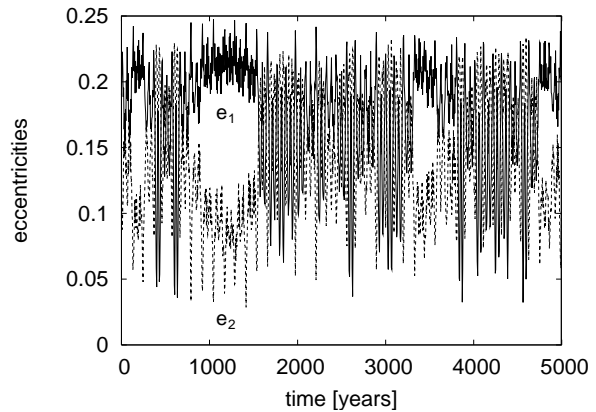} 
      \caption{Irregular behavior of the eccentricities of the giant planets in the system 
       HD~73526 using as initial conditions the data provided by the original dynamical 
       orbit fit (Table~\ref{table:1}). The upper (lower) curve shows the behavior of the  
       eccentricity of the inner (outer) planet.} 
         \label{Fig1} 
\end{figure}   
 
\begin{figure} 
   \centering 
   \includegraphics[width=8.5cm]{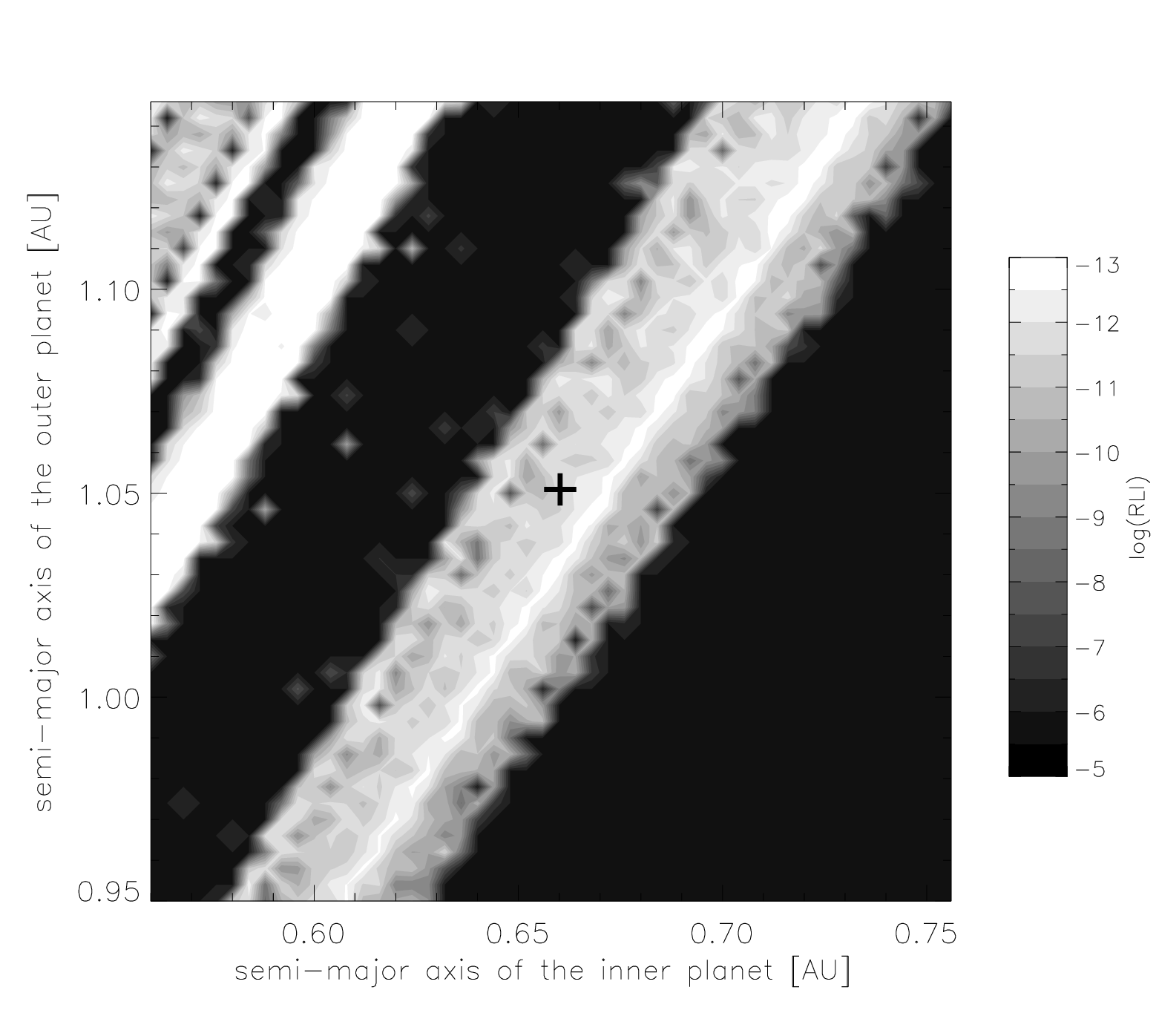} 
   \includegraphics[width=8.5cm]{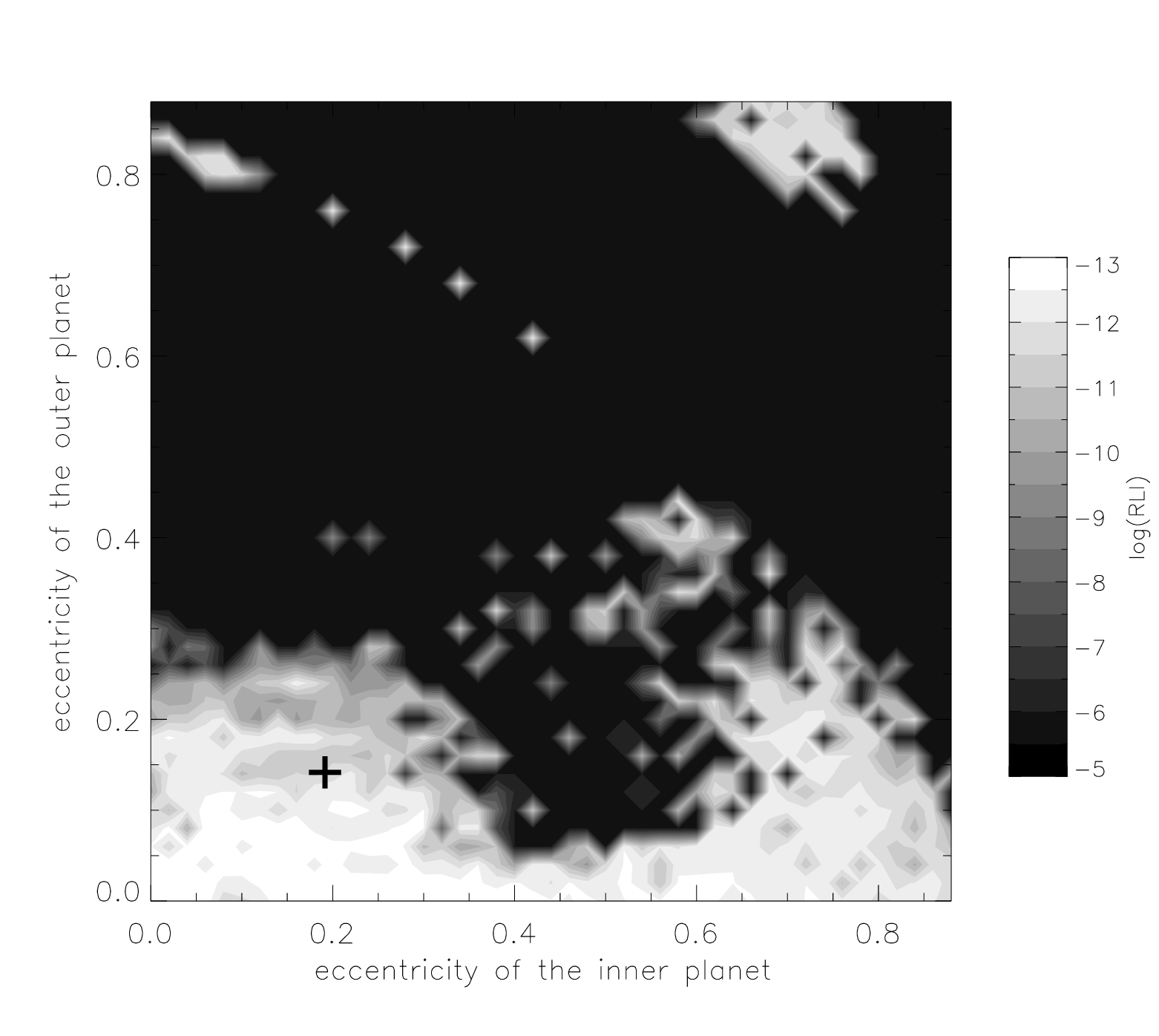} 
   \caption{Stability maps calculated by using the original dynamical orbit fit listed in 
   Table 1. The structure of the parameter planes indicates clearly that the orbital data 
   given by the original fit (marked by `$+$') are in a weakly chaotic region. Here brighter  
   areas refer to more stable and darker ones to unstable regions.} 
         \label{Fig2} 
\end{figure} 
 
\begin{table} 
\caption{Original dynamical orbit fit of HD~73526} 
\label{table:1}       
\centering                          
\begin{tabular}{c c c c c c}        
\hline\hline               
planet & mass [M$_J$]& $a$ [AU] & $e$ & $M$ [deg] & $\varpi$ [deg] \\    
\hline                       
   inner  & 2.9 & 0.66 & 0.19 &  86 & 203\\      
   outer  & 2.5 & 1.05 & 0.14 & 82 & 13 \\ 
\hline                                    
\end{tabular} 
\begin{tabular}{ c c c}        
\hline 
 $\chi_{\nu}^2$ & rms [m/s] & offset velocity [m/s] \\ 
\hline 
 1.57 &  7.9  & -29.96 \\ 
\hline\hline 
\end{tabular} 
\end{table} 
 
By using the Relative Lyapunov indicators \citep[RLI,][]{Sandoretal2004CeMDA}, 
we have mapped out the stability properties of the phase space in the close 
neighborhood of the original orbital data set, in a similar way done by 
\citet{Boisetal2003ApJ} in the case of HD~160691. 
We have calculated the stability properties of the $a_1-a_2$, $e_1-e_2$, $M_1-M_2$, 
and $\varpi_1-\varpi_2$ parameter planes, where $a$ is the semi-major axis, $e$ is 
the eccentricity, $M$ is the mean anomaly and $\varpi$ is the longitude of periastron 
of one of the giant planets, while the indices `1' and `2' refer to the inner and the 
outer planet, respectively.  
 
In Figure 2 we display the stability structures of the parameter planes for the 
semi-major axis and the eccentricities. During the calculation of a particular parameter 
plane the other orbital data have been kept fixed. This implies that when investigating 
the parameter plane of the semi-major axes $a_1-a_2$, the other orbital data such as 
the eccentricities ($e_1, e_2$), the mean anomalies ($M_1, M_2$), and the arguments 
of the periastrons ($\varpi_1, \varpi_2$) have been kept fixed to their original 
values (Table 1). On each parameter plane we have marked the stable regions by 
white, the weakly chaotic regions by grey and the strongly chaotic regions by black 
colors. The actual values of the corresponding orbital data are also shown by a 
symbol `$+$' on each parameter plane. By studying carefully the different stability 
maps (also the not displayed but calculated cases of the parameter planes 
$\varpi_1-\varpi_2$ and $M_1-M_2$), we conclude that the orbital data given 
by \citet{Tinneyetal2006ApJ} are embedded in a weakly chaotic region. We stress 
that this does not imply automatically the instability of their fit, however by using 
these orbital data the system exhibits irregular (or chaotic) behavior and may be 
destabilized at later times.  
 
We have also studied the stability of the system HD~73526 with slightly modified 
initial orbital elements. We have found, for instance, that even a small error of the initial 
semi-major axis ($\delta a_2=0.01$) or eccentricity ($\delta e_2=0.01$) of the outer 
planet leads to the destruction of the system in a few hundred thousand years. We 
note that the above errors ($\delta a_2$ and $\delta e_2$) are well inside the error 
limits of the original orbit fit of \citet{Tinneyetal2006ApJ}, being $\Delta a_2 = \pm 
0.08$ and $\Delta e_2 = \pm 0.09$.  Thus we believe that the use of initial orbital data 
resulting in regular orbits is more convenient when investigating the system 
HD~73526. 
 
On the other hand, the possibility that the system shows chaotic behavior as a result of 
its formation scenario cannot be excluded. It can also happen that a close encounter 
with a small mass body results in a chaotic behavior (characterized by the irregular 
variation of the eccentricities for instance), though the system itself remains stable for 
very long time. However, as our numerical simulations show, we have not found such 
a behavior; either the system shows regular behavior, or it is destroyed soon after the 
scattering event.  
 
\section{New orbital data and their stability} 
 
In order to find orbital solution for the system HD~73526 exhibiting regular behavior, 
we have used the Systemic Console (\texttt{http://www.oklo.org}), and have found 
some new orbital fits for the giant panets. In Table~2 we list four set of them, displaying 
different behavior. In the case of the first three sets of our orbital data, beside the mean 
motion resonance, the giant planets are also in apsidal corotation but with enlarged 
amplitudes in the resonant angles $\Theta_2$ or $\Delta \varpi$, while in the fourth 
case the apsidal corotation is no more present. In all four cases the eccentricities of the 
giant planets show relatively large oscillations. The behavior of the eccentricities is 
very similar to those found in the case of HD~128311 indicating clearly, that the 
present behavior of the system HD~73526 is not likely to be the result 
of a smooth adiabatic migration scenario alone.  
 
\begin{table}  
\caption{New dynamical orbit fits of the giant planets around HD~73526} 
\label{table:2}       
\centering                          
\begin{tabular}{c c c c c c c}        
\hline\hline 
Fit & planet & mass [M$_{\mathrm J}$]& $a$ [AU] & $e$ & $M$ [deg] & $\varpi$ [deg] \\    
\hline                       
 1 & inner  & 2.42 & 0.66 & 0.26 &  69.8 & 206.6\\      
    & outer  & 2.58 & 1.045 & 0.16 & 163.2 & 265.6 \\ 
\hline 
 2 & inner  & 2.62 & 0.66 & 0.209 &  77.7 & 208.5\\      
    & outer  & 2.56 & 1.047 & 0.194 & 131 & 316.2 \\  
\hline 
  3 & inner  & 2.415 & 0.659 & 0.26 &  70.7 & 202.9\\      
     & outer  & 2.55 & 1.045 & 0.107 & 170.7 & 253.7 \\  
\hline    
  4 & inner  & 2.675 & 0.66 & 0.209 &  80.1 & 207.1\\      
     & outer & 2.53 & 1.048& 0.172 & 122.4 & 327.6 \\ 
\hline\hline 
\end{tabular} 
\end{table} 
 
\begin{table}[] 
\caption{Properties of the dynamical orbit fits for HD~73526.} 
\label{table:3}       
\centering                          
\begin{tabular}{c c c c}        
\hline\hline 
Fit No. & $\chi_{\nu}^2$ & rms [m/s] & offset velocity [m/s] \\ 
\hline 
1 & 1.81 &  8.36 & -46.67 \\ 
2 & 1.60 & 8.09 & -33.76 \\ 
3 & 1.87 & 8.4 & -49.54 \\ 
4 & 1.58 & 8.04 & -32.33 \\ 
\hline\hline 
\end{tabular} 
\end{table} 
 
\begin{figure} 
   \centering 
   \includegraphics[width=8.5cm]{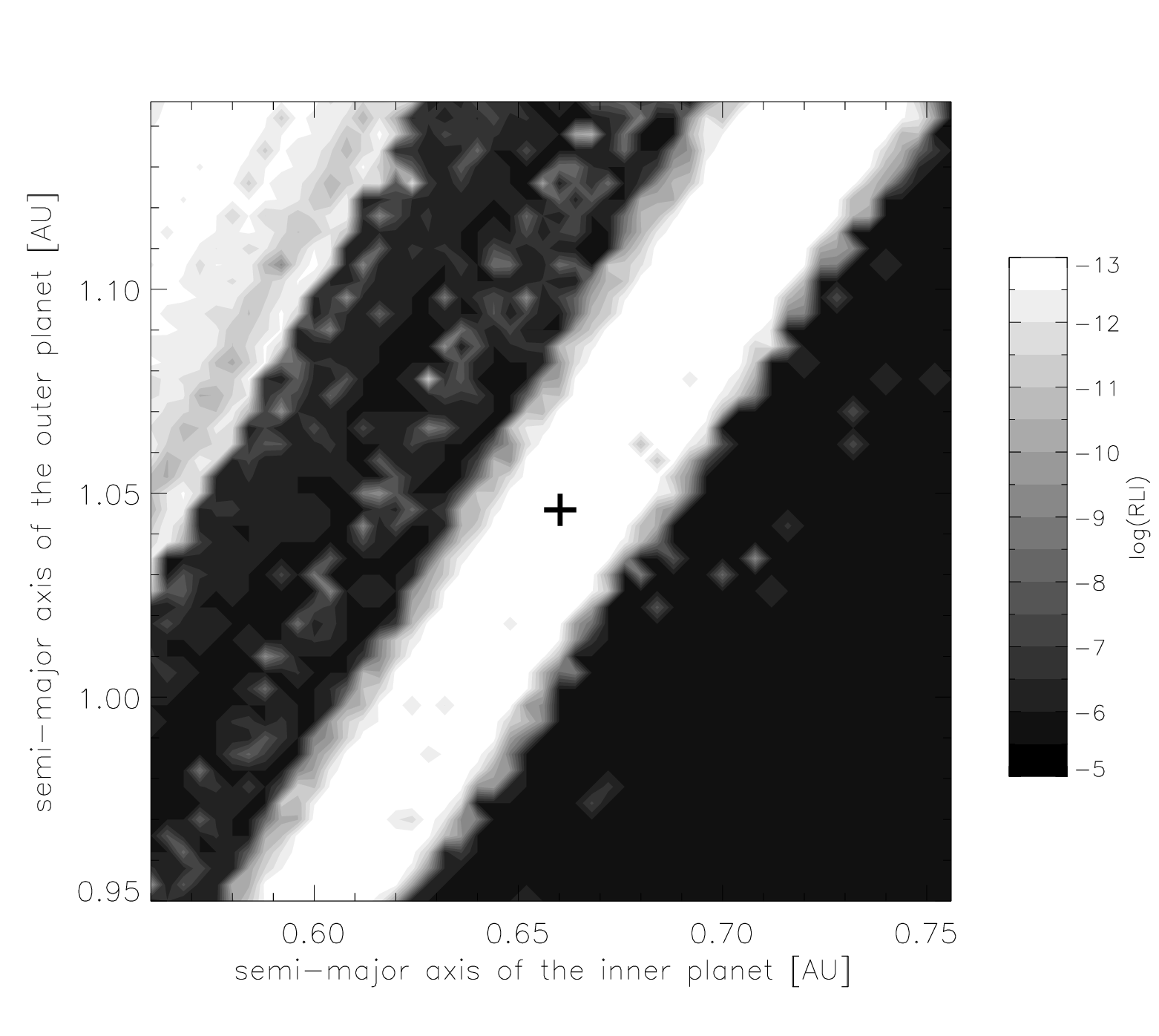} 
   \includegraphics[width=8.5cm]{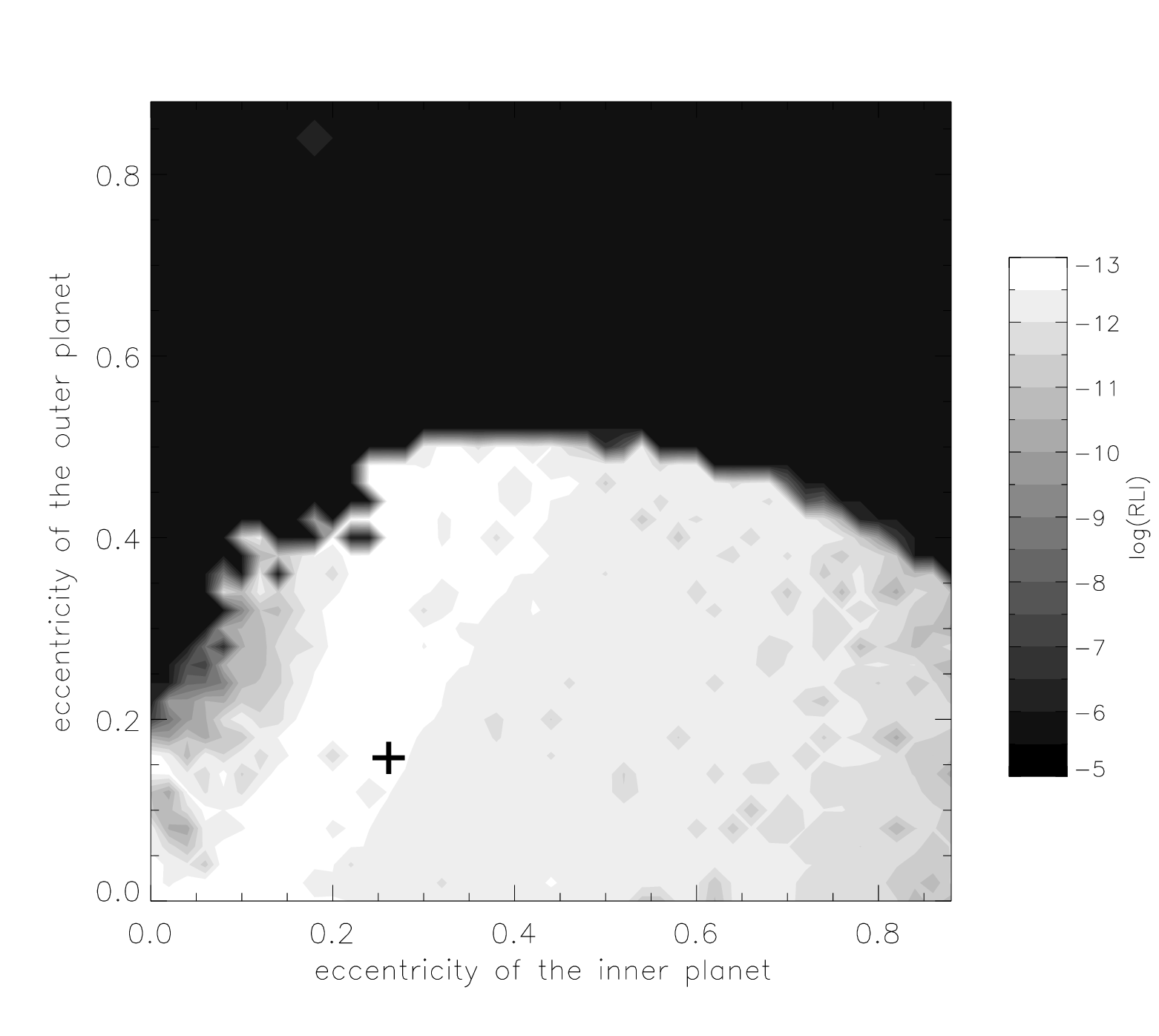} 
    \caption{Stability maps calculated by using the orbital data provided by Fit 1 of 
    Table 2. It can be seen from the parameter planes $a_1-a_2$ ({\bf top}) and 
    $e_1-e_1$ ({\bf bottom})  that the orbital data of Fit 1 (marked by `$+$') are 
    embedded well in the stable white region.} 
         \label{Fig3} 
\end{figure} 
 
In order to investigate the stability of the newly derived orbital elements (shown in 
Table 2), we have computed (by using the Relative Lyapunov indicator) a series of 
stability maps, similar to those shown in Figure 2. We have found that in all cases the 
orbital elements are deeply embedded in the stable regions of the parameter planes 
$a_1-a_2$, $e_1-e_2$, $\varpi_1-\varpi_2$, and $M_1-M_2$. In Figure 3 we 
show the stability structure of the parameter planes $a_1-a_2$ and $e_1-e_2$ in the 
case of Fit 1. The semi-major axes and eccentricities corresponding to Fit 1 are 
marked by the symbol `$+$', and they almost lie in the middle of the stability 
region (white regions of the parameter planes). We note that at first sight the structure 
of the parameter planes $a_1-a_2$ corresponding to the original orbital data (shown in 
Figure 2) and to the orbital data of Fit 1 (shown in Figure 3) are quite similar to each 
other. However, in Figure 2 the stable regions are very narrow, the light regions 
(which may indicate stability) are mainly grey and have a fuzzy structure indicating 
clearly the weakly chaotic character of the system. In Figure 3 the light regions are 
white and they have very homogeneous structure showing ordered and therefore 
stable behavior of the system.  
 
We have also checked by numerical integration whether the orbital data shown in 
Table 2 result in a reliable radial velocity curves indicating the motion of the star. Two 
examples are shown in Figure 4, corresponding to the Fit 1 and Fit 4, respectively. 
We note that the radial velocity curves of Fit 1 and Fit 3 are very similar to each other, 
while the radial velocity curves of Fit 2 and Fit 4 are very similar to the radial velocity 
curve given by \citet{Tinneyetal2006ApJ} corresponding to their original dynamical 
orbit fit. In the case of Fit 1 and Fit 3 the mass of the inner giant planet is (slightly) 
smaller than the mass of the outer planet, which is also supported by the planetary 
migration scenario. According to full hydrodynamical simulations a cavity opens 
between the giant planets, and the inner planet orbits in a low density gaseous  
environment. The outer giant planet is still connected to the disk thus it can accrete 
more material, which may result in larger mass for it than for the inner planet 
\citep{Kleyetal2004A&A}. 
 
\begin{figure} 
   \centering 
   \includegraphics[width=8cm]{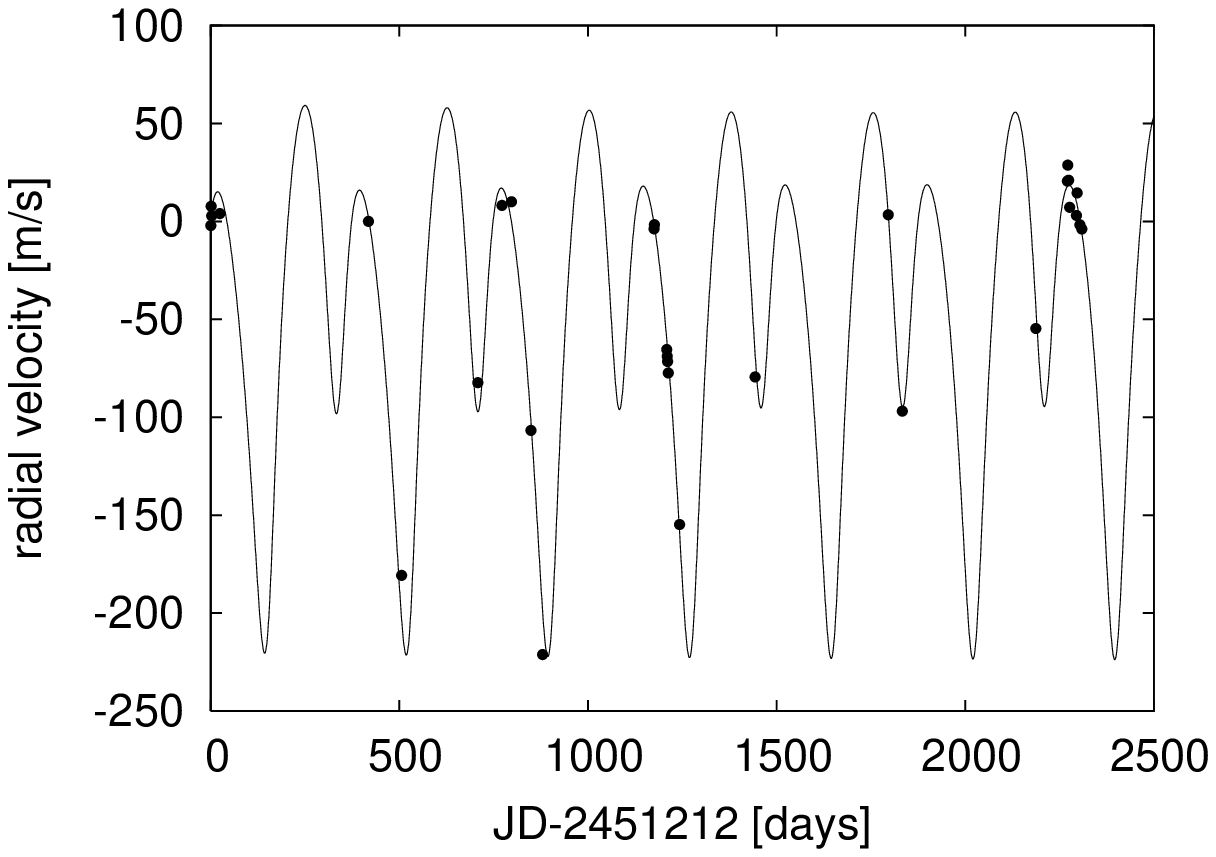} 
   \centering 
   \includegraphics[width=8cm]{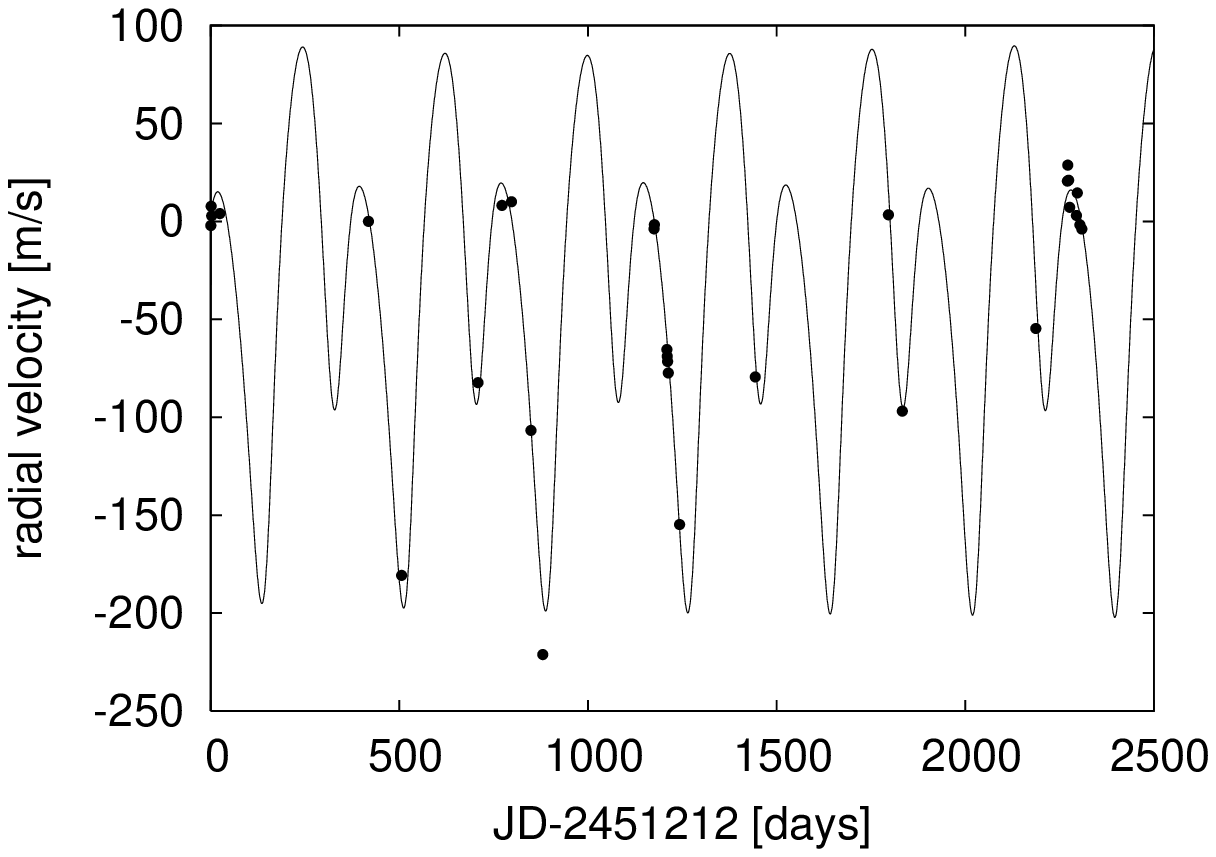} 
   \caption{Radial velocity curves of the star obtained by numerical integration using 
   as initial conditions the orbital data given by Fit 1 (top panel) and Fit 4 (bottom 
   banel) of Table 2. It can be seen that the calculated curves fit well to the measured 
   radial velocity points (black dots).} 
         \label{Fig4} 
\end{figure}   
 
\begin{figure} 
   \centering 
   \includegraphics[width=8cm]{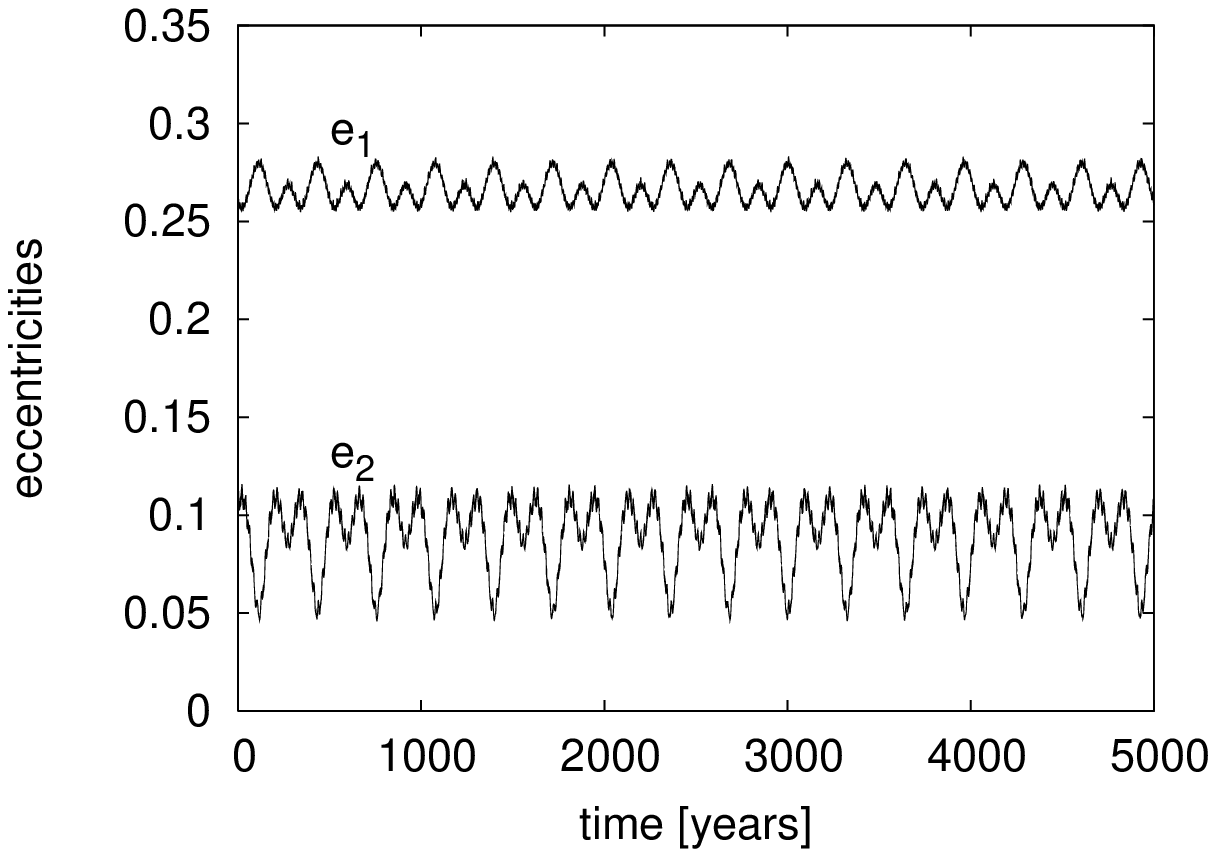} 
   \centering 
   \includegraphics[width=8cm]{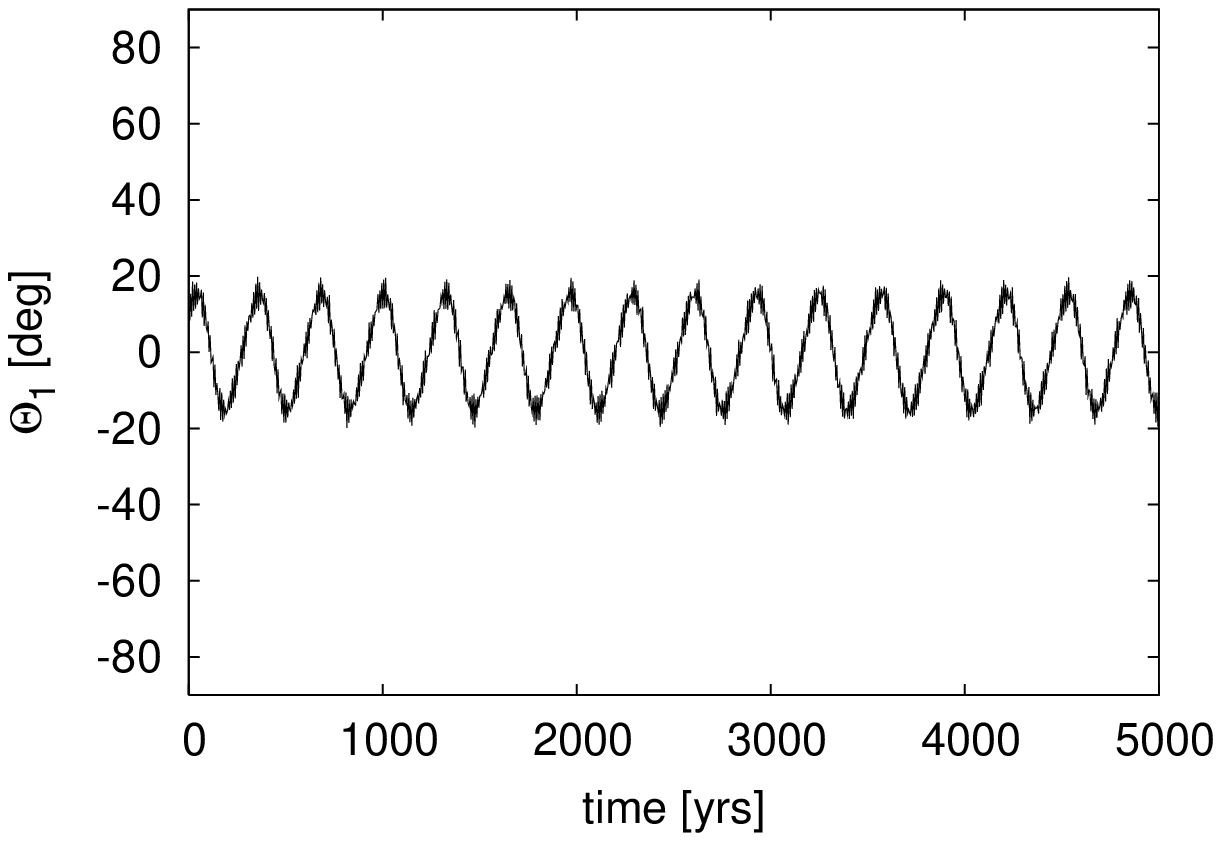} 
   \centering 
   \includegraphics[width=8cm]{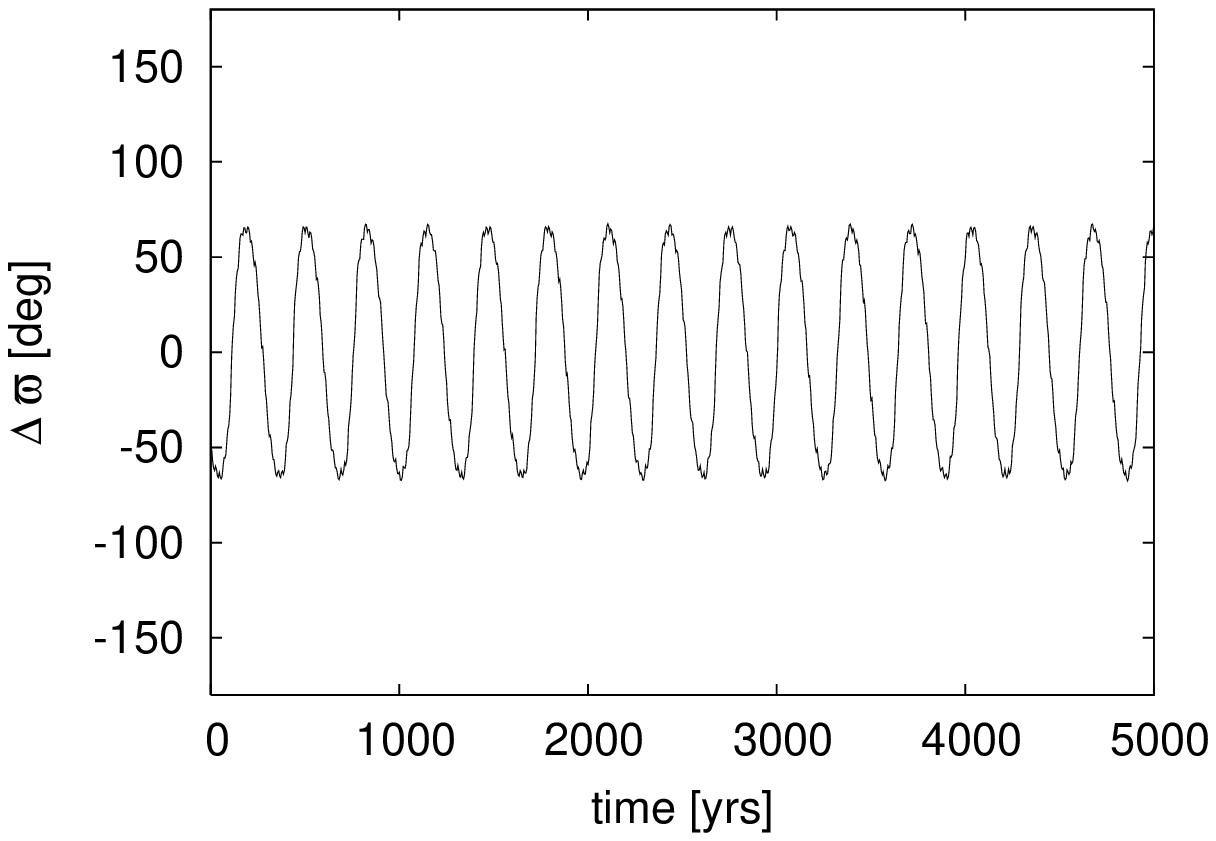} 
   \caption{The behavior of the eccentricities $e_1$ and $e_2$ ({\bf top}), the 
   resonant angles $\Theta_1$ ({\bf middle}) and $\Delta\varpi$ ({\bf bottom}) of 
   the giant planets by using the initial orbital data given by Fit 3.} 
         \label{Fig5} 
\end{figure}  
 
The similarities between Fit 1 and Fit 3 or Fit 2 and Fit 4 can also be observed in the 
behavior of the eccentricities of the giant planets.  The eccentricities in the cases of Fit 
1 and 3 show moderate oscillations, while in the cases of Fit 2 and 4 they oscillate 
with very large amplitudes.  
One example for the behavior of the eccentricities is shown in the top of Figures \ref{Fig5} 
corresponding to Fit 3. The other two (middle and bottom) panels of Figure \ref{Fig5} show 
the behavior of the resonant angles $\Theta_1$ and $\Delta\varpi$, respectively, 
corresponding again to Fit 3.   
 
In the following, when investigating the formation of the system HD~73526, we 
intend to model a similar behavior provided by Fit 1 and Fit 3, in which case the outer 
planet is a little bit more massive than the inner one, and the oscillations in the 
eccentricities are more moderate than in the cases of Fit 2 and Fit 4. 
 
\section{Formation of the system HD~73526 by smooth inward migration} 
 
Since the present behavior of the planetary system around HD~73526 is very similar 
to the behavior of the system HD~128311, we shall investigate in this section the 
possibility whether this behavior could be the result of a mixed formation scenario 
combining a smooth migration with a sudden perturbative effect as suggested by 
\citet{Sandor&Kley2006A&A} in the case of HD~128311. We first model the 
evolution of the system by applying a slow inward migration. Having estimated  
the basic characteristics of the planetary migration, we intend to study the formation  
of the system taking into account (i) a fast dispersal of the protoplanetary disk 
(resulting in the termination of the migration of the outer giant planet) and (ii)  
planet-planet scattering. 
 
\subsection{A hydrodynamical model for HD~73526} 
 
To study the general feasibility of forming a resonant system such as HD~73526 
through a migration process of two embedded planets we performed full 
hydrodynamic evolutions where the accretion disk is treated in a flat two-dimensional 
approximation. The general setup and the applied numerical methods of the models 
are identical to those used in the detailed study of GJ~876 by 
\citet{Kleyetal2005A&A}, see also \cite{Kley1999MNRAS}.  
Here we restrict ourselves to a description of the particulars of our model for 
HD~73526, and concentrate of new features. 
 
For the masses of the two planets we use $m_1=2.415 M_J$ and 
$m_2=2.55 M_J$ (see Fit~3 from Table~2), where $M_J$ is the mass of Jupiter,  
and the planets are not allowed to accrete any mass during their evolution. 
The planetary orbits are initially circular at distances of $r_1 =1$ and $r_2 = 2$AU 
from the central star with mass $1.08 M_\odot$. The flat disk extends from  
of $r_{min} = 0.2$ to $r_{max} =4.0$AU with an initial surface density profile of 
$\Sigma (r) = \Sigma_0 r^{-1/2}$.  
Here, the density at $r_0=1$AU is normalized such that $\Sigma_0 r_0^2 / M_\odot = 
7.22 \cdot 10^{-4}$ which gives a total disk mass in the domain of about $0.024 
M_\odot$. To make the initial evolution of the planets more realistic we superimpose 
initial density gaps to the unperturbed $r^{-1/2}$-profile \citep{Kley1999MNRAS, 
Kley2000MNRAS}. The disk is driven by an $\alpha$-type viscosity with $\alpha = 
0.01$, and the temperature is specified using a fixed relative vertical thickness of $H/r 
 = 0.05$, no energy equation is solved. The smoothing length for the gravitational 
potential is $\epsilon = 0.6 H$, which gives approximately the correct values of the 
planetary migration rate when compared to the full three-dimensional case.  
For the torque cutoff, i.e. the region around the planet which is excluded in the 
calculation of the torques, we use $r_{torq} = 0.6 R_{Roche}$.  
In the regular (non-resonant) evolution of the planets the exact values of 
$\epsilon$ and $r_{torq}$ should not influence the results too strongly due to the 
lack of material in the extended gap regions surrounding the planets. When the 
eccentricities of the planets increase during the evolution they periodically move 
through the disk at their peri- and apo-center and these numerical parameter may 
become more important. We have checked that our results do not depended on the 
exact values of $\epsilon$ and $r_{torq}$ as long as they are in the above specified 
range.  
 
The computational domain from $r_{min}$ to $r_{max}$ is covered by 240 radial and  
504 azimuthal gridcells where the radial spacing is logarithmic and the azimuthal  
equidistant. This gives roughly squared gridcells throughout the whole domain.  
At the outer boundary we use damping boundary conditions where the disk profiles  
are relaxed to the initial profile \citep{2006MNRAS.370..529D}. 
At the inner boundary we use a new type of boundary condition where 
we specify a outflow condition with an azimuthally averaged radial velocity which  
has the magnitude of  
\begin{equation} 
\label{eq:vrad} 
 v_r (r_{min}) = - 5 \, \,  \frac{3}{2} \, \frac{\nu}{r_{min}}, 
\end{equation} 
The typical viscous radial inflow velocity of accretion disks is  
$v_{visc}(r) = 3 \nu /(2 r)$. 
We use here a 5 times higher value than  $v_{visc}(r_{min})$ to accomodate 
an increased clearing of the inner disk due to {\it i}) the disturbances of the 
embedded planets and {\it ii}) the vicinity of the central star. 
Nevertheless, this boundary condition will prevent material leaving the inner disk 
too rapidly which is typically observed in hydrodynamic evolutions with 
embedded planet \citep{2007A&A...461.1173C}. 
 
\begin{figure} 
   \centering 
    \includegraphics[width=0.95\linewidth]{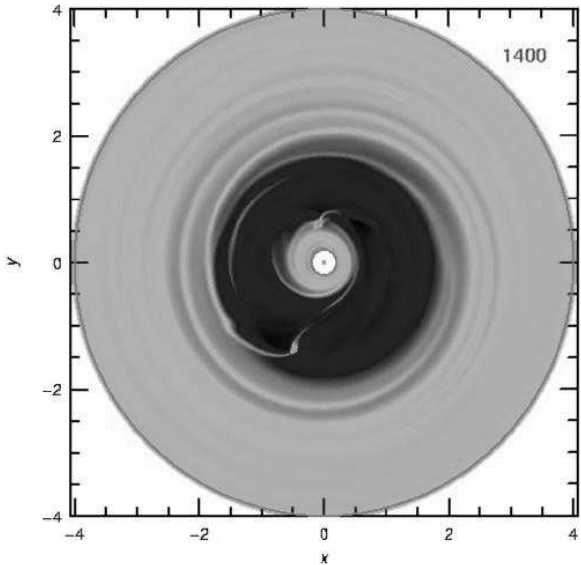} 
    \caption{Two dimensional density distribution of a protoplanetary accretion disk 
                  with two embedded planets after about 1400 years.  From their initial 
		   positions ($a_1=1.0$ and $a_2=2.0$~AU) the planets have migrated 
		   inwards and, after capturing into a 2:1 resonance, have reached 
		   $a_1=0.85$ and $a_2=1.36$~AU with eccentricities  
		   $e_1=0.32$ and $e_2=0.08$. Within the inner planet the inner disk has 
		   not been cleared even after such a long  evolution time due to a modified 
		   new inner boundary condition for the radial inflow velocity.} 
     \label{fig:u1450-2d} 
\end{figure}

In Fig.~\ref{fig:u1450-2d} we show the distribution of the surface density 
$\Sigma(r,\varphi$) of the protoplanetary accretion disk after about 1400 years.   
The inner planet is located at $x=-0.11, y=-0.61$ and the outer one at $x=-0.48, 
y=-1.37$. In contrast to previous models where the two planets orbit in an inner cavity 
of the disk without any inner disk \citep{Kleyetal2004A&A}, here the inner disk has 
not been cleared and is still present due to the more realistic inner boundary condition 
given in Eq.~(\ref{eq:vrad}). 
 
\begin{figure} 
   \centering 
    \includegraphics[width=0.95\linewidth]{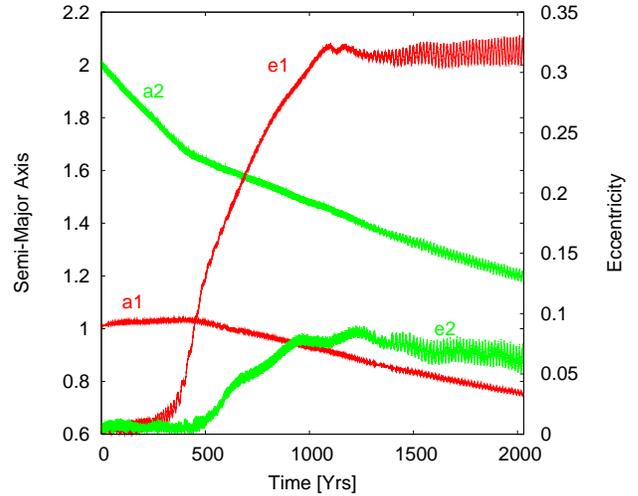} 
    \caption{The evolution of the semi-major axis and eccentricities of the two 
  embedded planets for the full hydrodynamical model. The inner planet 
  ($a_1, e_1$) 
 is indicated by the darker curves. After 400 yrs the outer planet captures 
 the inner on in a 2:1 mean motion resonance and they are coupled during  
 their inward migration. 
 The eccentricities increase rapidly after resonant capture and settle to equilibrium 
  values for larger times.} 
         \label{fig:ae-c03} 
\end{figure}

As a consequence the inner planet still is in contact with the inner disk which exerts 
gravitational torques on it. These tend to push the planet outward and will produce 
a damping of the eccentricities. 
In Fig.~\ref{fig:ae-c03} we display the evolution of eccentricities and 
semi-major axis of the two planets. Initially, both planets orbit in a wide joint gap 
where the outer planet experiences only the torques of the outer disk and migrates 
inward, while the inner disk pushes the inner planet slightly outwards. 
Capture in the 2:1 mean motion resonance occurs at 400~yrs after which the planets 
migrate jointly 
towards the star. The eccentricities increase strongly and reach equilibrium values 
at approximately 1100~yrs. In contrast additional full hydrodynamical test models that do not 
have an inner disk show a continued eccentricity increase in particular for the outer 
planet. We attribute this reduced magnitudes of the eccentricities to the damping 
action of the inner disk. When the eccentricity of the inner planet increases it will 
periodically have to enter into the inner disk and the gravitational 
torques will tend to reduce its eccentricity. Equilibrium is then given by the 
eccentricity driving caused by the inward motion of the planets and the damping 
caused by the disk. After resonant capture the resonant angles $\Theta_1$ and $\Delta 
\varpi$ are both librating with a libration amplitude of $ \approx 35^\circ$ for 
$\Theta_1$ and $< 10^\circ$ for $\Delta \varpi$. 
Hence, this hydrodynamical evolution displays an adiabatic migration process which 
results in apsidal corotation of the two planets, where the apsidal lines are always 
aligned.  
 

In the long term the eccentricity of the outer planet ($e_2$) remains at a level 
of 0.05 to 0.07 and does not decline any further. Additional test computations 
using different resolutions and boundary conditions confirm this trend.  
The shown hydrodynamical simulation represents an explorative study to 
demonstrate that such a damping mechanism of the inner disk may have operated. 
Using this mechanism it might be possible to accomodate a larger radial migration 
for the resonant planets than that found previously for the case of GJ~876 
\citep{Kleyetal2005A&A} because eccentricties will not be pumped up to 
very large values. 
This topic will be investigated in more detail in future work. 
 
\begin{figure} 
   \centering 
   \includegraphics[width=8cm]{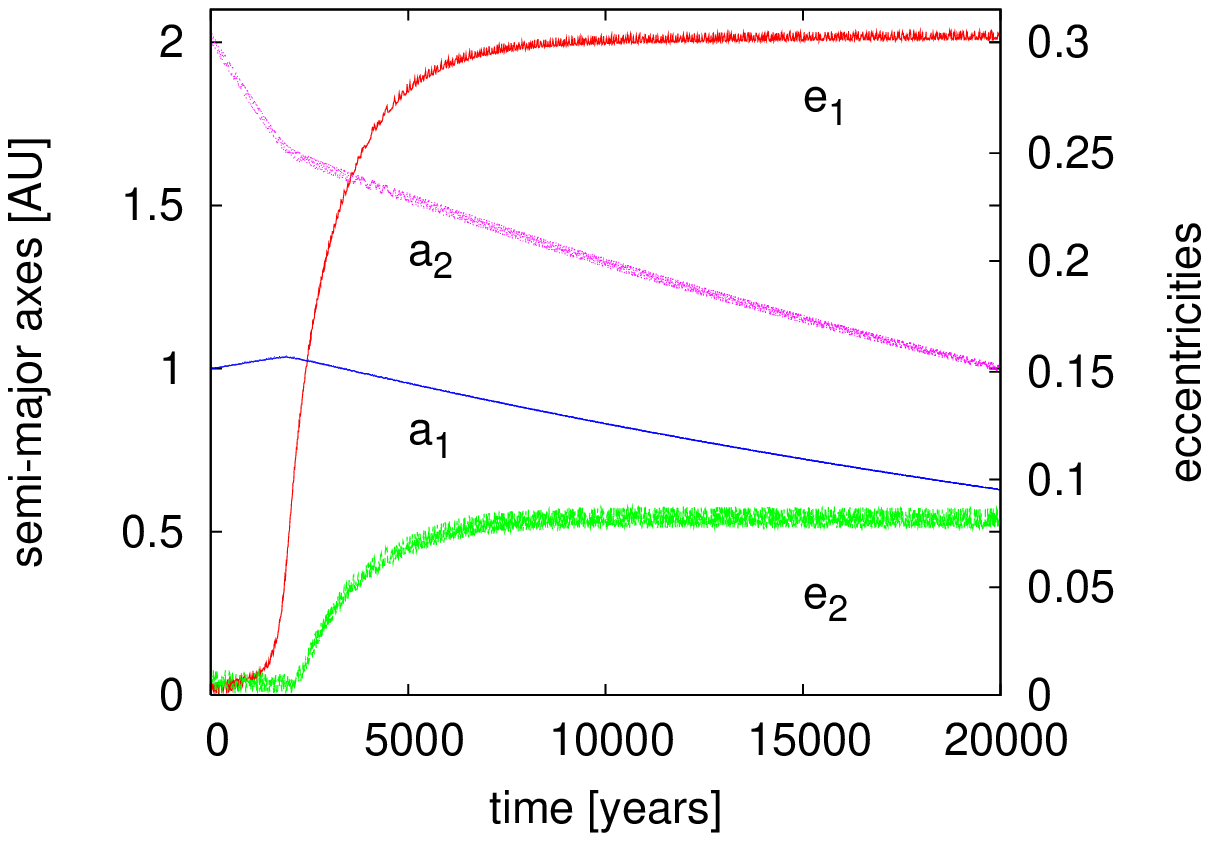} 
   \includegraphics[width=8cm]{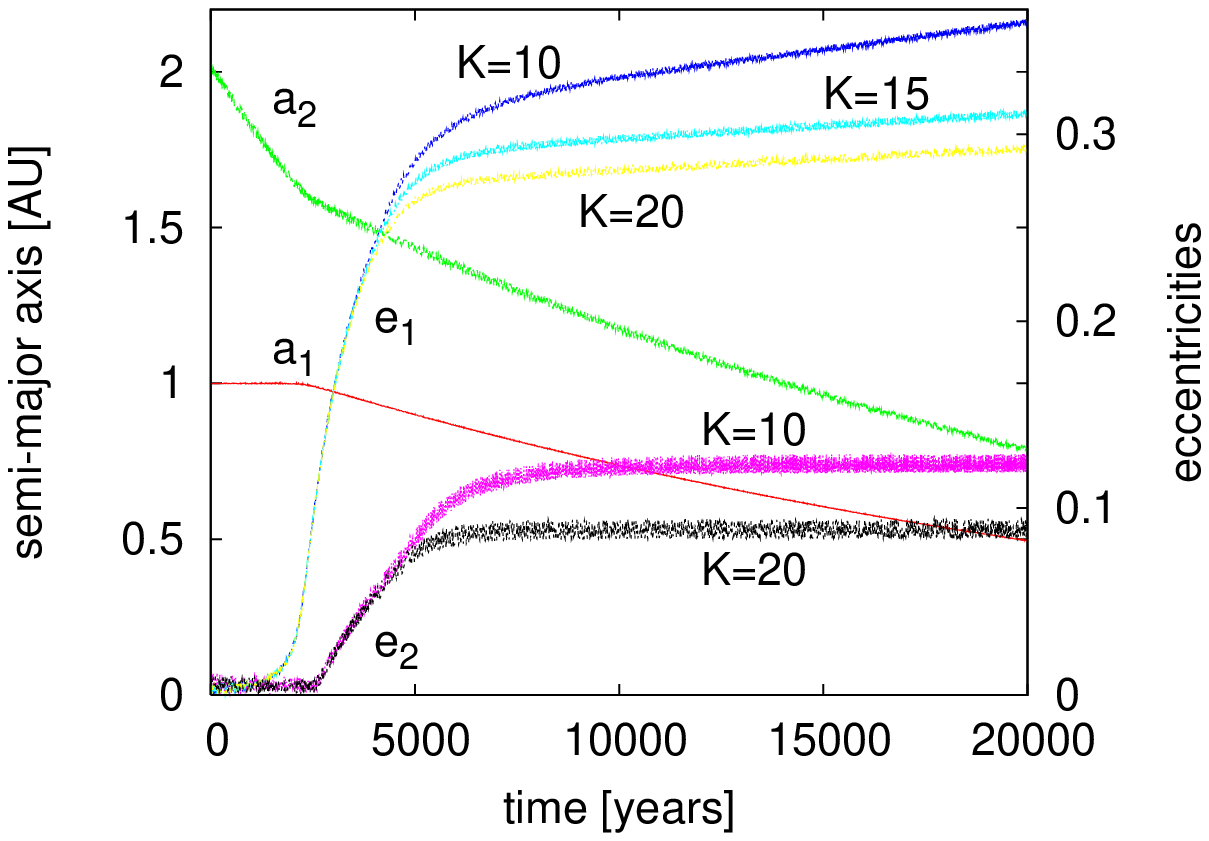} 
   \includegraphics[width=8cm]{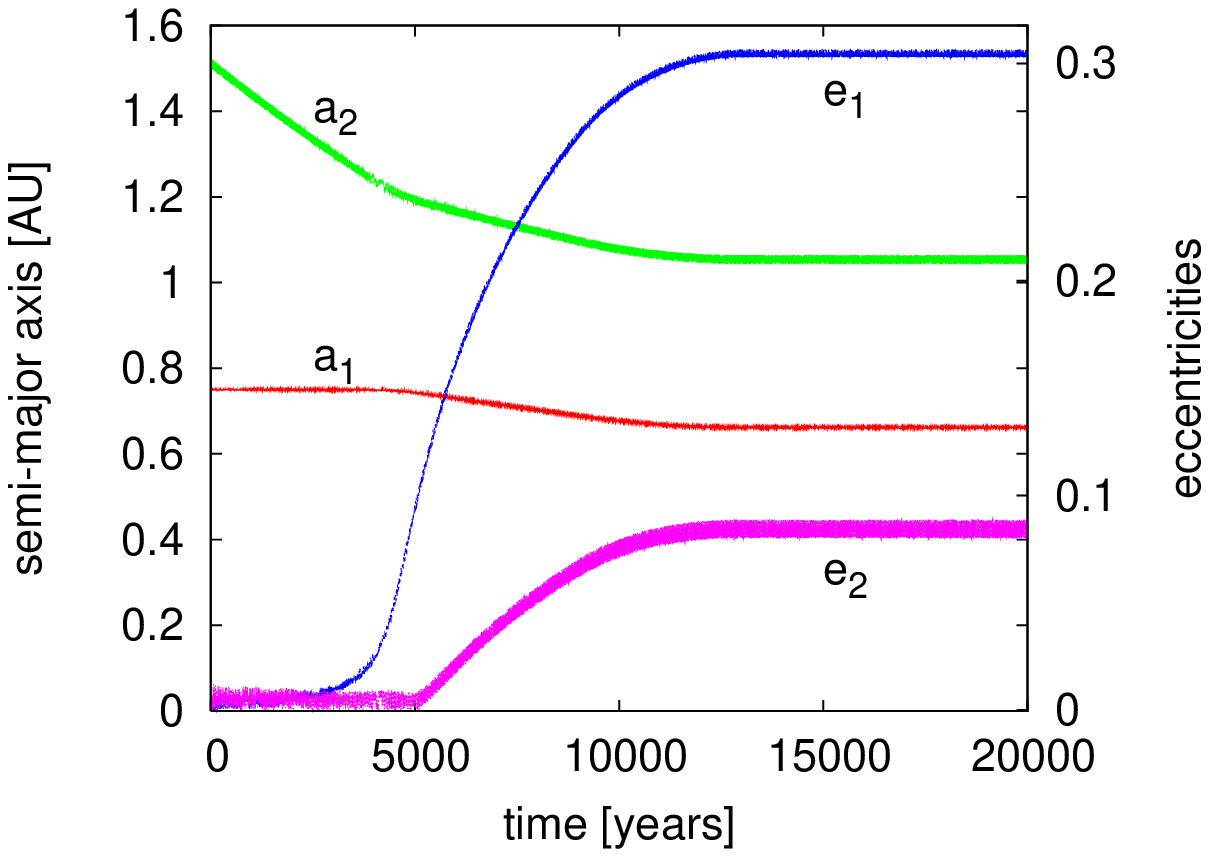} 
    \caption{Evolution of the semi-major axes and the eccentricities of the giant planets 
     during different migration scenarios: (i) the actions of an outer and an inner disk are 
     modeled, $\tau_{a_2}=10^4$ years, $\tau_{a_1}=-5\times 10^4$ years and 
     $K_1=K_2=10$ ({\bf Top});  (ii) only the damping effect of an outer disk is modeled, 
     $\tau_{a_2}=10^4$ years and $K=10$, $15$, and $20$ ({\bf Middle}); and (iii)  
     the resonant capture takes place right before the disk's dispersal obeying a linear 
     reduction law between $9\times 10^3$ and $1.3\times 10^4$ years, 
     $\tau_{a_2}=2\times 10^4$ years, and $K=10$ ({\bf Bottom}).} 
         \label{fig:migrscen} 
\end{figure} 
\begin{figure} 
   \centering 
   \includegraphics[width=8cm]{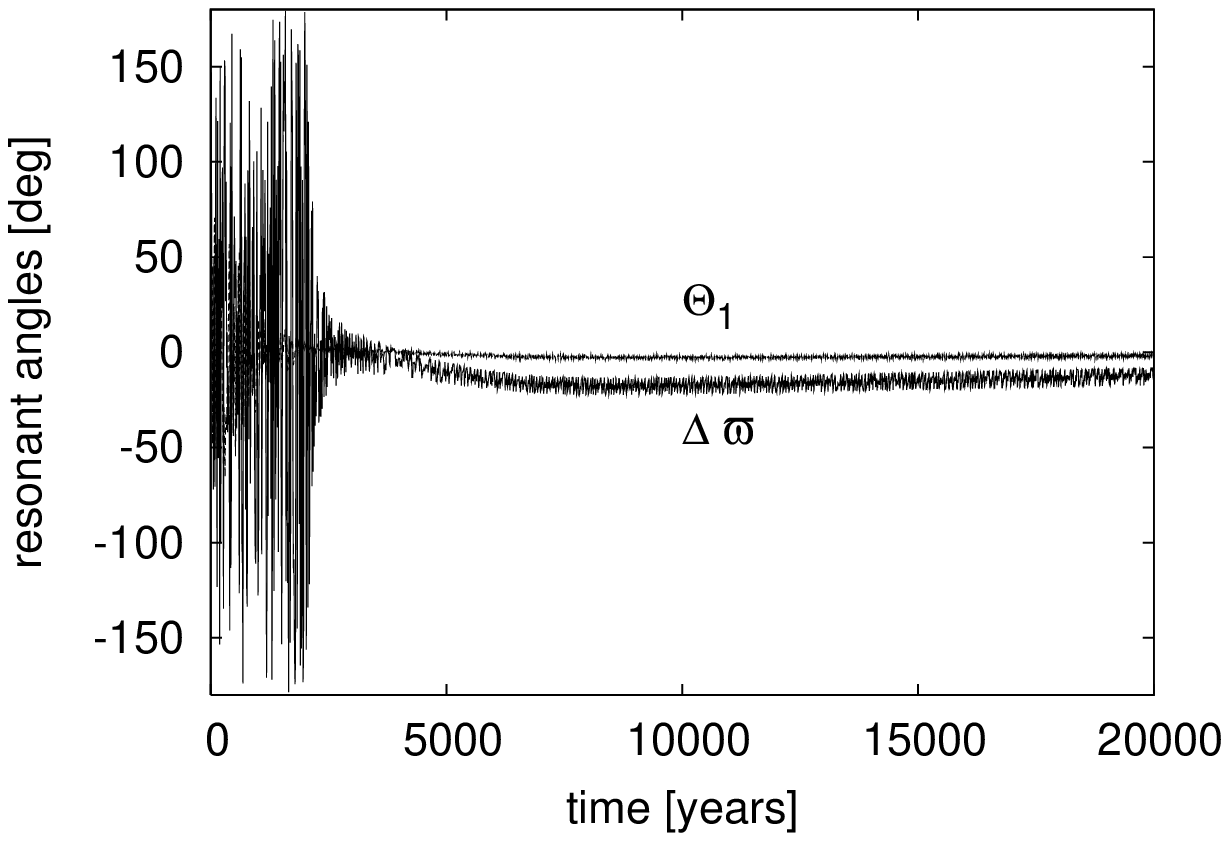} 
    \includegraphics[width=8cm]{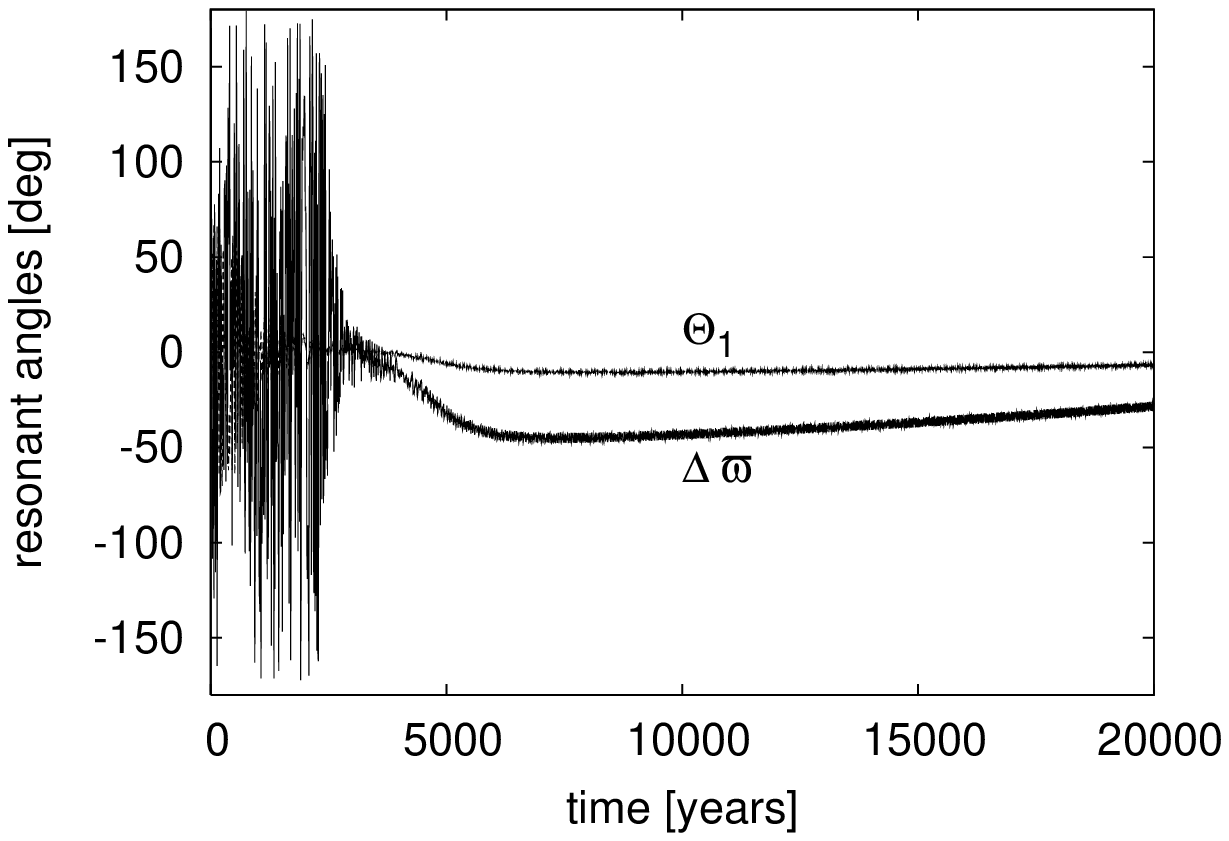} 
    \includegraphics[width=8cm]{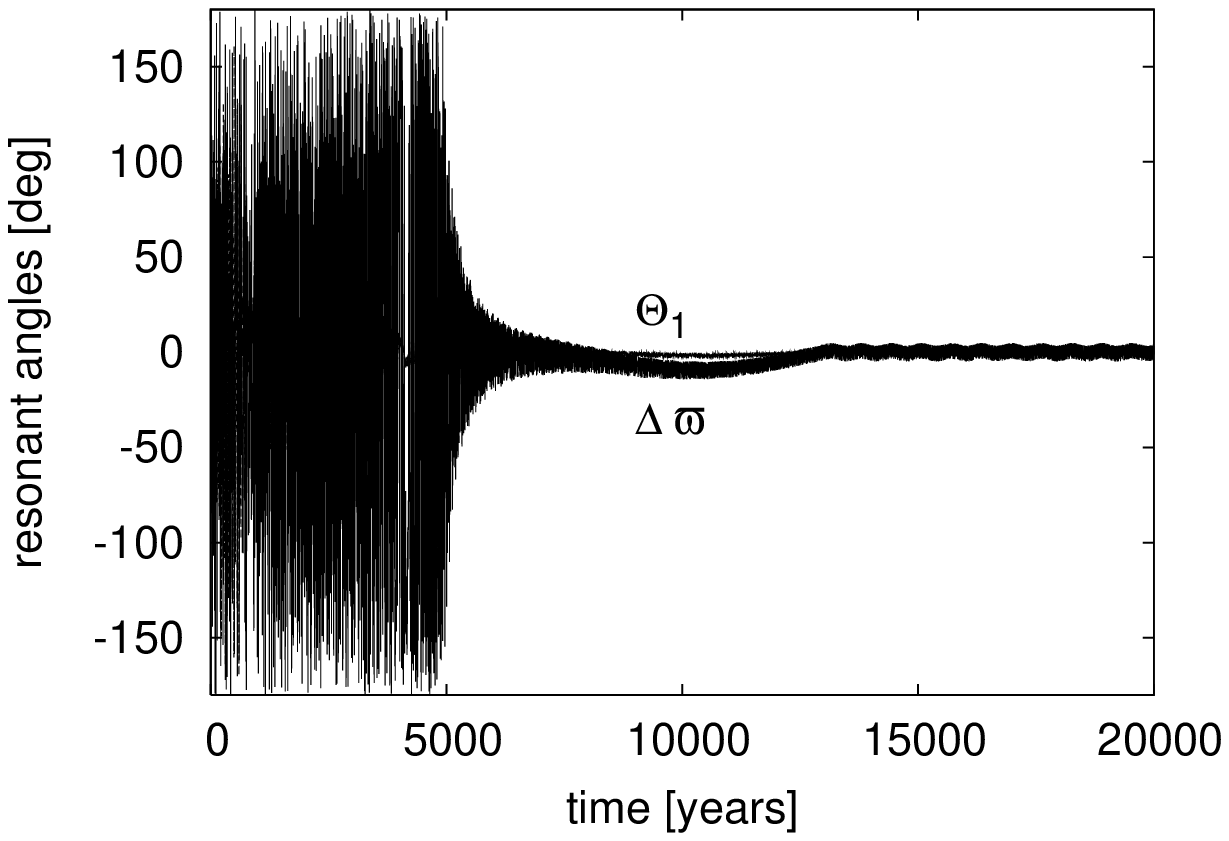} 
   \caption{Evolution of the resonant angles $\Theta_1$ and $\Delta\varpi$ during  
   the migration scenarios corresponding those displayed in Figure \ref{fig:migrscen}.}  
   \label{fig:resangmigrscen} 
\end{figure}

\subsection{Slow migration modeled by dissipative forces} 
 
Hydrodynamical calculations modeling the migration of planets embedded in a 
gaseous protoplanetary disk require relatively large computational time. On the other 
hand, the damping effect of a protoplanetary disk to the orbital evolution of the 
embedded giant planets can also be modeled conveniently in the framework of the 
gravitational N-body problem by using well chosen non-conservative forces. These 
additional forces can be parametrized by the migration rate $\dot a/a$ and the 
eccentricity damping rate $\dot e/e$, or by the corresponding $e$-folding times 
$\tau_a$ and $\tau_e$ of the semi-major axis and eccentricity of the outer planet, 
respectively \citep[see][]{Lee&Peale2002ApJ,Beaugeetal2006MNRAS}. The 
relations between the damping rates and $e$-folding times are $\dot a/a = - 1/\tau_a$, 
and similarly for the eccentricities $\dot e/e = - 1/\tau_e$. We define the ratio 
between the $e$-folding times $K=\tau_a/\tau_e$, (or $\dot e/e = - K |\dot a/a|$), 
which according to \citet{Lee&Peale2002ApJ}, determines the final state of the 
system in the case of a sufficiently slow migration. 
 
We recall that in the migration scenarios used up to now mainly the damping of the 
outer giant planet has been taken into account. As we have mentioned previously, we 
have also supposed that beside the outer disk there is also an inner disk, and the 
planets migrates in a cavity between these disks. The presence of the inner disk 
accelerates the inner giant planet, thus forces it to migrate outward and also damps its 
eccentricity. This effect can also be modeled by using a (repelling) non-conservative 
force also parametrized by $\dot a/a$ (having positive sign for outward migration), 
and $K$ or $\dot e/e$. 
 
In what follows, we present our results obtained in modeling the formation of the system 
HD~73526 by a slow migration process. We have studied three cases. In the first case, 
beside the damping effect of the outer disk, we have taken into account the effect of an 
inner disk. In the second case we have assumed that only the outer planet is affected 
by the outer disk and forced to migrate inward. Finally, in the third case we have 
assumed that after the resonant capture the protoplanetary disk dissapears gradually 
inhibiting the further increase of the inner planet's eccentricity. In this latter case only 
the outer planet has been forced to migrate inward. Our aim is to provide an evolution 
of the system which is in accordance with the eccentricity limits given by Fits 1-4.    
 
In order to compare the results of dissipative N-body calculations to those obtained by 
hydrodynamical simulations, at the beginning of the migration the giant planets move 
on circular orbits at distances $a_1=1$AU and $a_2=2$AU around a 
$M_*=1.08M_{\odot}$ mass star. The masses of the giant planets have been fixed to 
$m_1=2.415M_{\mathrm J}$ and $m_2=2.55M_{\mathrm J}$, where  
$M_{\mathrm J}$ is the mass of Jupiter. For the inward migration of the outer giant 
planet we have 
applied $\tau_{a_2}=10^4$ years, while for the outward migration of the inner giant 
planet $\tau_{a_1}=-5\times 10^4$ years. The ratio between the e-folding times in 
both cases was $K=10$. Comparing Figure \ref{fig:ae-c03} with the top panel of 
Figure \ref{fig:migrscen}, one can conclude that the presence of an inner disk can be 
modeled by a properly chosen dissipative force. The behavior of the resonant angles 
for this case is shown in the top panel of Figure \ref{fig:resangmigrscen}. Initially 
both $\theta_1$ and $\Delta\varpi$ deviate slightly from $0^{\circ}$ and tend finally 
to $0^{\circ}$. N-body calculations also confirm the results of the hydrodynamical 
simulations that an inner disk can damp efficiently the eccentricity of an inner planet 
during a permanent migration.  
 
Hydrodynamical simulations for HD~73526 also show that the lack of the inner disk 
results in a continuous increase of the eccentricity of the inner giant planet. In what 
follows, we will study which value of ratio $K$ is necessary to stop the increase of the 
inner planet's eccentricity when no inner disk is considered. (It is known from the 
hydrodynamical simulations that $K$ lies typically between $1-10$.)   
 
We have applied a damping of the semi-major axis of the outer planet 
with an $e$-folding time 
$\tau_{a_2}=10^4$ year using different $K$ values ($K=10$, $15$, and $20$). The 
time evolution of the semi-major axes and the eccentricities are shown in the middle 
panel of Figures \ref{fig:migrscen}, while the behavior of the resonant angles 
$\theta_1$ and $\Delta\varpi$ (with $K=10$) is displayed also in the middle panel of 
Figures \ref{fig:resangmigrscen}. Studying the behavior of the eccentricities for the 
different ratios $K$ one can see that during the migration process the eccentricity of 
the outer planet is damped sufficiently, while the eccentricity of the inner planet is 
slightly increasing. If the migration is terminated when the semi-major axes reach their 
actual values (at $t\approx 1.3\times 10^4$ year), we need $K\sim 15$ not to exceed 
the observed eccentricity of the inner planet. (The upper limit for the eccentricity of the 
inner planet is around $e_{1\mathrm{max}}\approx 0.3$, as it has been shown by our 
numerical integrations based on the orbital data of Fits 1-4.) On the other hand, from 
hydrodynamical simulations we know that $K\sim 1-10$, thus although the above 
values $K=15$ seem to be a bit high, it may be acceptable for thick 
disks. 
But the $K$ value can be reduced to $K_1=K_2=10$ by the presence of an 
inner disk 
as shown above by the hydrodynamic simulations. The 
evolution of the resonant angles $\Theta_1$ and $\Delta\varpi$ (for $K=10$) are 
shown in the middle a panel of Figure \ref{fig:resangmigrscen}. After the resonant 
capture, $\Theta_1$ and $\Delta\varpi$ deviate from $0^{\circ}$, and having 
reached their extrema ($\sim 20^{\circ}$ for $\Theta_1$ and $\sim 50^{\circ}$ for 
$\Delta\varpi$), they tend to $0^{\circ}$ very slowly. The reason of this behavior 
might be that the mass ratio between the giant planets ($m_1/m_2=0.947$) lies very 
close to the critical mass ratio $(m_1/m_2)_{\mathrm{crit}}=0.95$ given by 
\citet{Lee2004ApJ}. If $m_1/m_2>(m_1/m_2)_{\mathrm{crit}}$, the resonant 
system evolves temporarily through an asymmetric apsidal configuration. We note that 
the deviation of the angles $\Theta_1$ and $\Delta\varpi$ is the largest in the the 
above case when only the outer giant planet is damped by an outer disk.  
 
It is also clear that the migration of the giant planets should terminate when they reach 
their actual positions. If the termination of the migration is a slow process the system 
reaches a state very close to a periodic solution of the corresponding three-body 
problem studied by \citet{Psychoyos&Hadjidemetriou2005CeMDA}, for instance.  
 
In order to avoid the use of the relatively high values for $K\sim 15 - 20$, we have 
also studied the possibility when the resonant capture between the giant planets takes 
place just before the dispersal of the protoplanetary disk. We recall that this scenario 
\citep[proposed by][]{Kleyetal2005A&A} might have occurred in the case of 
GJ~876, if there is not acting a strong damping mechanism on the eccentricities. We 
have supposed that prior to the resonant capture the giant planets migrated separately to 
their orbits having $a_1=0.75$ AU, and $a_2=1.5$ AU. Then the outer planet 
migrates inward with an e-folding time of its semi-major axis $\tau_a=2\times 10^4$ yr 
with $K=10$, and captures the inner planet into the $2:1$ resonance, see the bottom 
panel of Figure \ref{fig:migrscen}. The resonant capture takes place around $t\approx 
5\times 10^3$ years, the eccentricity of the inner planet grows rapidly, but before 
exceeding the limit $e_{1\mathrm{max}} = 0.3$, the protoplanetary disk has already 
disappeared and thus the migration is terminated. The disk dispersal happens between 
$9\times 10^3 - 1.3\times 10^4$ years, obeying a linear reduction law. During this 
simulation the final value of the inner planet's eccentricity remained around the critical 
value $e_1 \leq 0.3$, however the ratio of the $e$-folding times was only $K=10$.  
 
\begin{figure} 
   \centering 
   \includegraphics[width=8cm]{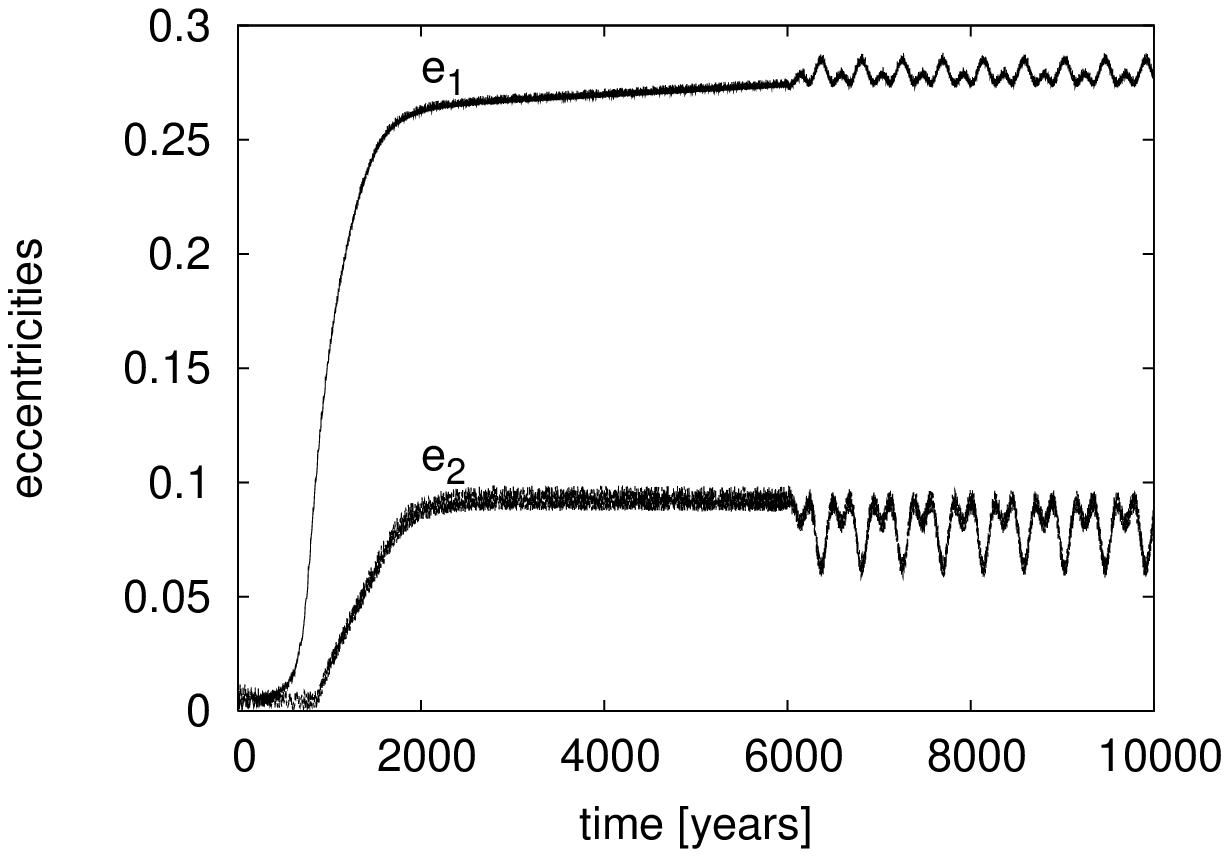} 
   \includegraphics[width=8cm]{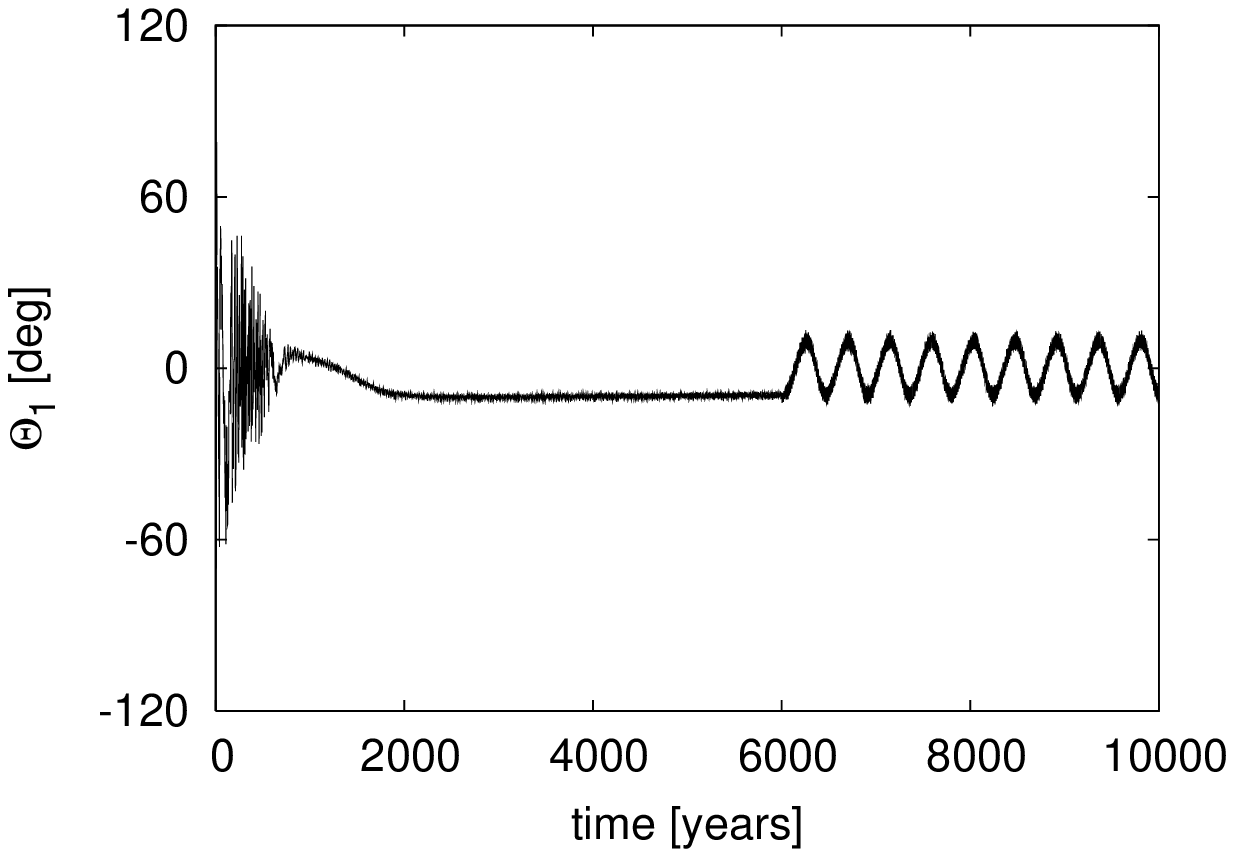} 
   \includegraphics[width=8cm]{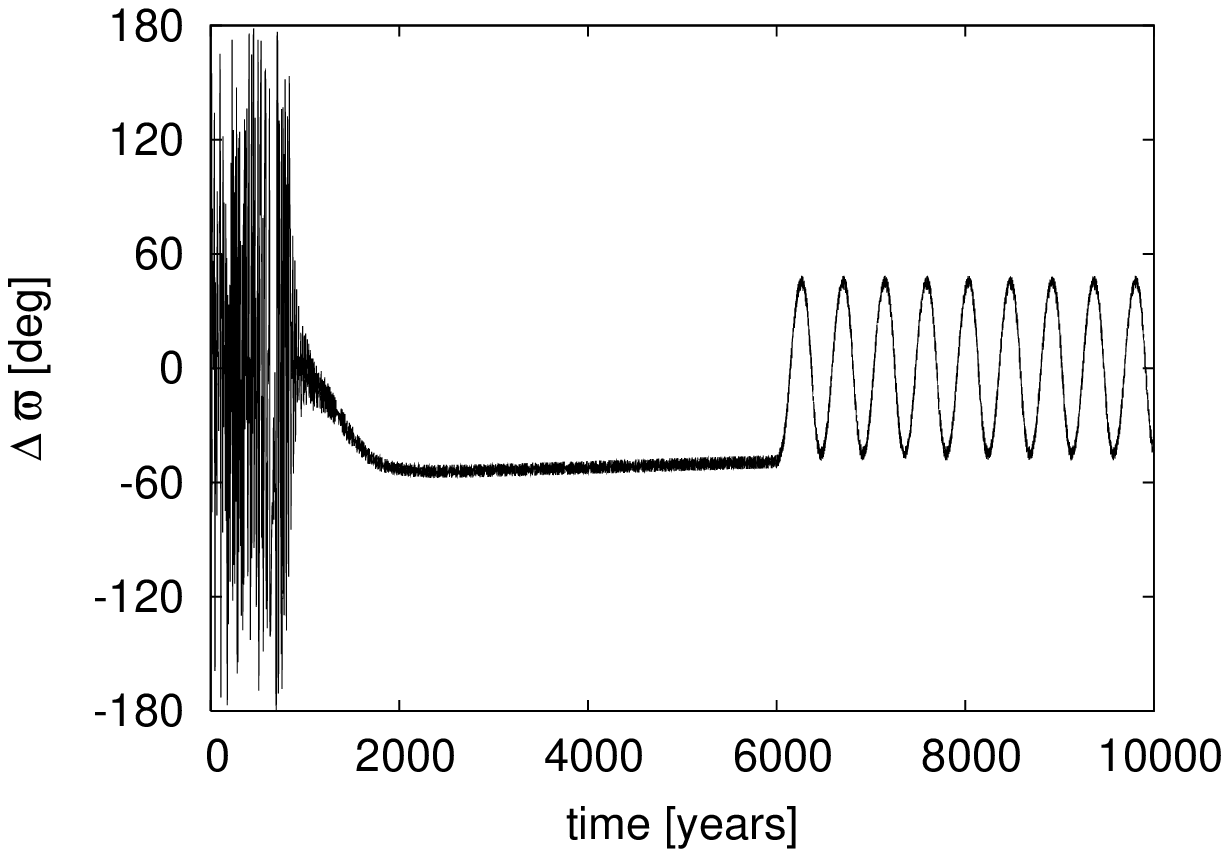} 
   \caption{{\bf Top}: Behavior of the eccentricities of the giant planets during  
   an inward migration followed a sudden dispersal of the protoplanetary disk, 
   $\tau_{a_2}=4\times 10^3$ and $K=15$. The giant planets orbit initially at $a_1=1$ 
   AU, $a_2=2$ AU, and the migration is stopped between $6000$ and $6050$ yrs, 
   applying a linear disk dispersal law. The fast termination of the migration results in 
   oscillations of the eccentricities of the planets, very similar to that shown in Figure 4. 
   {\bf Middle} and {\bf Bottom}: Behavior of the resonant angles $\Theta_1$ and 
   $\Delta\varpi$ corresponding the above scenario. After the fast termination of the 
   migration the resonant angles also behave very similar to the case displayed in  
   Figure 4.}
   \label{fig:sdterm} 
\end{figure} 
 
\section{Formation of the system HD~73526 by mixed evolutionary scenarios} 
 
In the previous section we have presented three possible migration scenarios that can push 
the giant planets deep into the $2:1$ resonance not contradicting to the observed upper 
limits of the eccentricities ($e_1\sim 0.3$, $e_2\sim 0.1$ ). On the other hand, as our 
dynamic orbit fits (Fit 1-4) in Table~\ref{table:2} show, the eccentricities of the  
giant planets and the resonant 
angles oscillate with considerably amplitude. This behavior is similar to that one has 
been observed in the case of the system HD~128311, thus in what follows, we intend 
to model the formation and early evolution of the system HD~73526 by using mixed 
evolutionary scenarios melting together inward migration and sudden perturbative 
events mentioned in \citep{Sandor&Kley2006A&A}.   
 
\subsection{Sudden termination of the migration} 
 
Recent {\it Spitzer} observations of young stars show that the inner part of the 
protoplanetary disk may be emptied due to photoevaporation induced by the central star 
\citep[see][]{DAlessioetal2005ApJ,Calvetetal2005ApJ}. Thus when approaching the 
inner rim of such a disk, the inward migration can be stopped rapidly 
\citep{2006ApJ...642..478M}.  

If the inner rim is located at a given radius of the 
disk (e.g. it is stationary) the migration cannot be maintained because in this case the 
the outer disk, which is the driving agent of the migration, cannot follow the planet. 
on its inward drift.
 
Based on our experience in modelling the formation of the system HD~128311, we have also 
investigated the above case when the termination of the migration (due to the  
presence of a stationary inner rim) happens in a very short timescale, here  
$\Delta t_{\mathrm dis}=50$ years. In this case we apply a faster migration 
$\tau_a=4\times 10^3$ years and $K=15$. The behavior of the eccentricities and the 
resonant angles $\Theta_1$ and $\Delta\varpi$ is shown in Figure \ref{fig:sdterm}. It 
is clear that this fast termination of the migration can result in a behavior of the 
eccentricities and the resonant angles, which is similar to the observed case shown in 
Figure 5.  
 
The reason of this behaviour is that in the beginning of the migration process, the 
resonant angles deviate from $0^{\circ}$, reach their extrema, and tend very slowly to 
$0^{\circ}$ (see also the middle panel of Figures \ref{fig:resangmigrscen}). If the 
termination of the migration occurs when the resonant angles still deviate from 
$0^{\circ}$, they begin to oscillate (with an amplitude equal to their values at the end 
of the disk's dispersal) around their equilibrium (which is in this case $0^{\circ}$). This 
scenario works also in the cases of slower migration rates, however in the case of a 
faster migration the deviation of the resonant angles from 
$0^{\circ}$ is larger, and hence the amplitude of the oscillations in $e$, is noticeably 
higher.   
 
We should note that the above fast migration of protoplanets may not be likely in 
protoplanetary disks. In thick disks fast migration may occur, but in this case is not 
easy to find relevant physical processes leading to the sudden termination of migration. 
However, the outcome of the sudden stop of migration characterized with the above 
parameters yields results being in good agreement with the behavior of the system 
by using as initial conditions the data of Fit 3.    
 
\begin{figure} 
  \centering 
  \includegraphics[width=8cm]{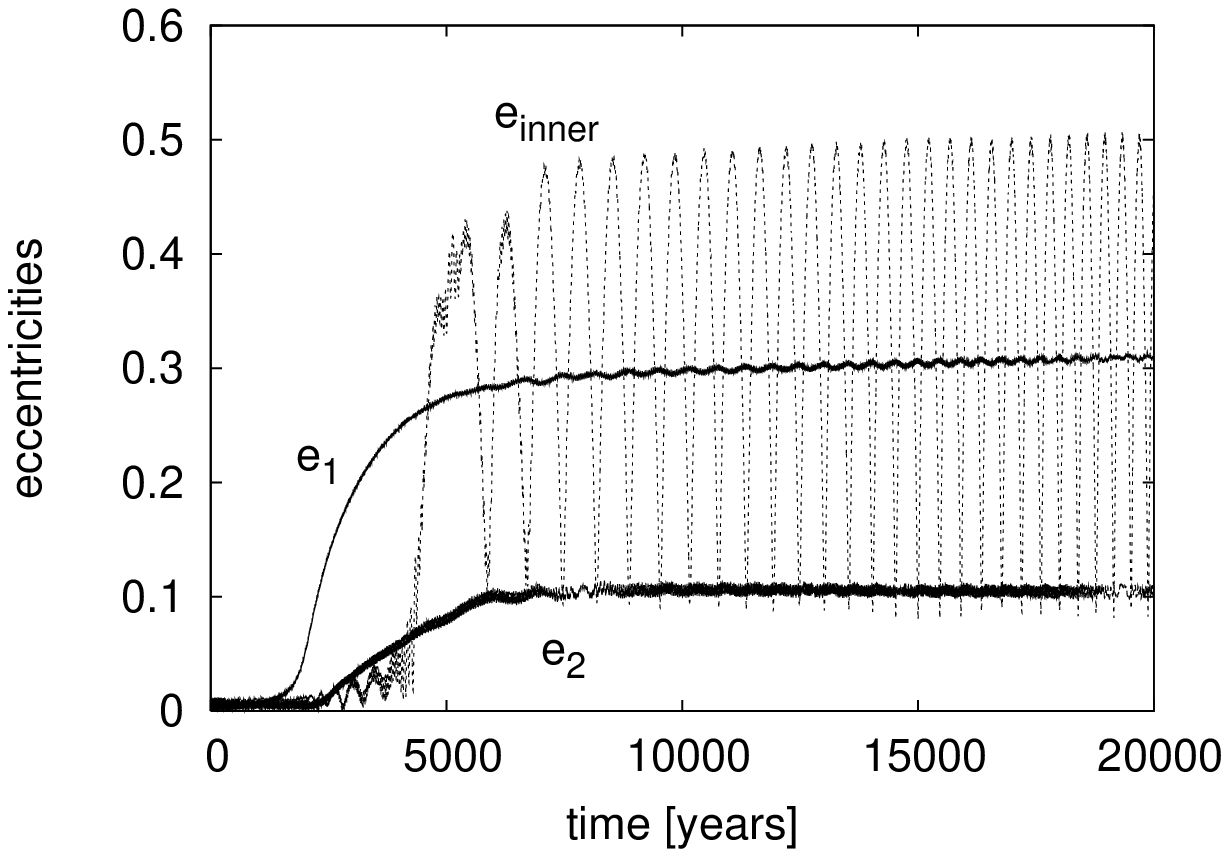} 
  \centering 
  \includegraphics[width=8cm]{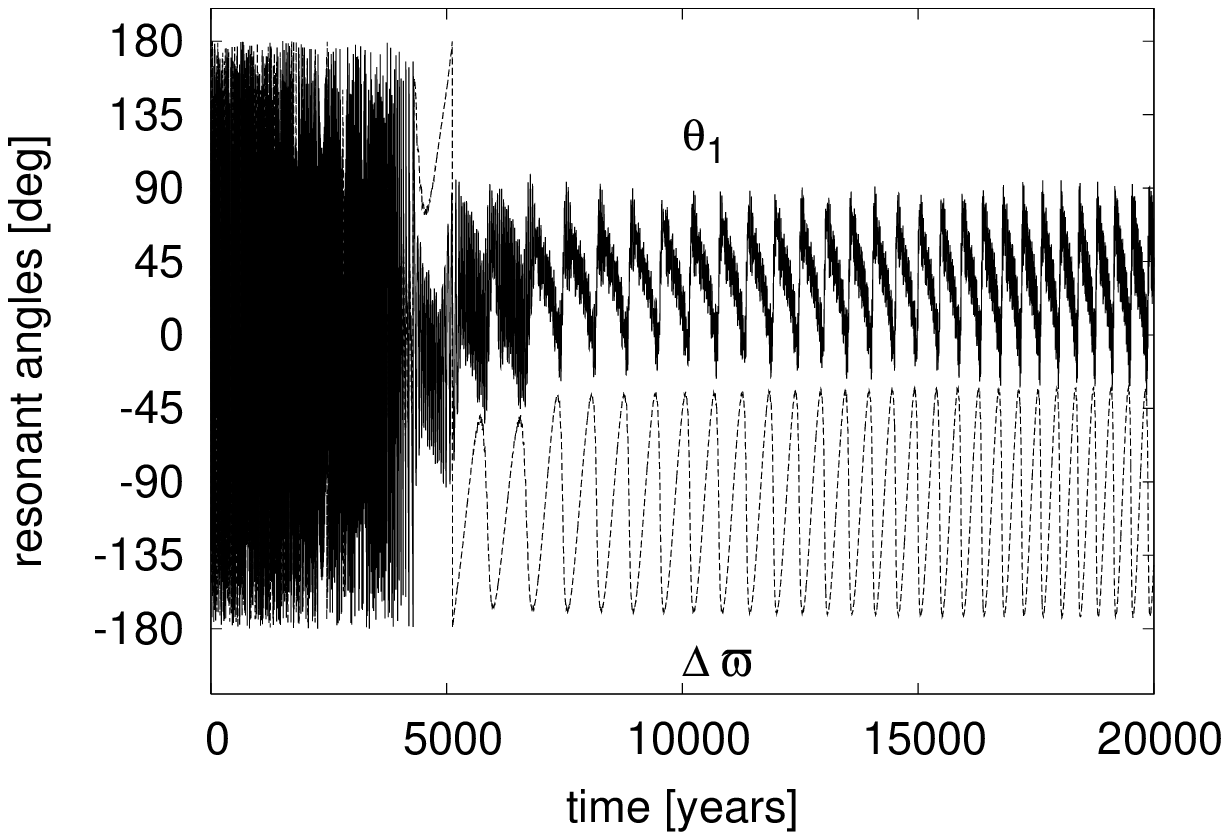} 
  \caption{{\bf Top}: The evolution of the eccentricities of the giant and the small  
  mass planets during a planetary migration when the inner giant planet captures the 
  small mass inner planet into a protective $3:1$ resonance. This resonant configuration 
  inhibits the increase of the small mass inner planet's eccentricity thus the system 
  remains stable under the whole migration process. {\bf Bottom}: The behavior of the 
  resonant angles $\theta_1$ and $\Delta\varpi$ shows that beside the $3:1$ 
  resonance, the orbits of the inner giant and the small mass planet also are in an 
  asymmetric apsidal corotation.} 
         \label{fig:capt31prot} 
\end{figure} 
 
\begin{figure} 
  \centering 
  \includegraphics[width=8cm]{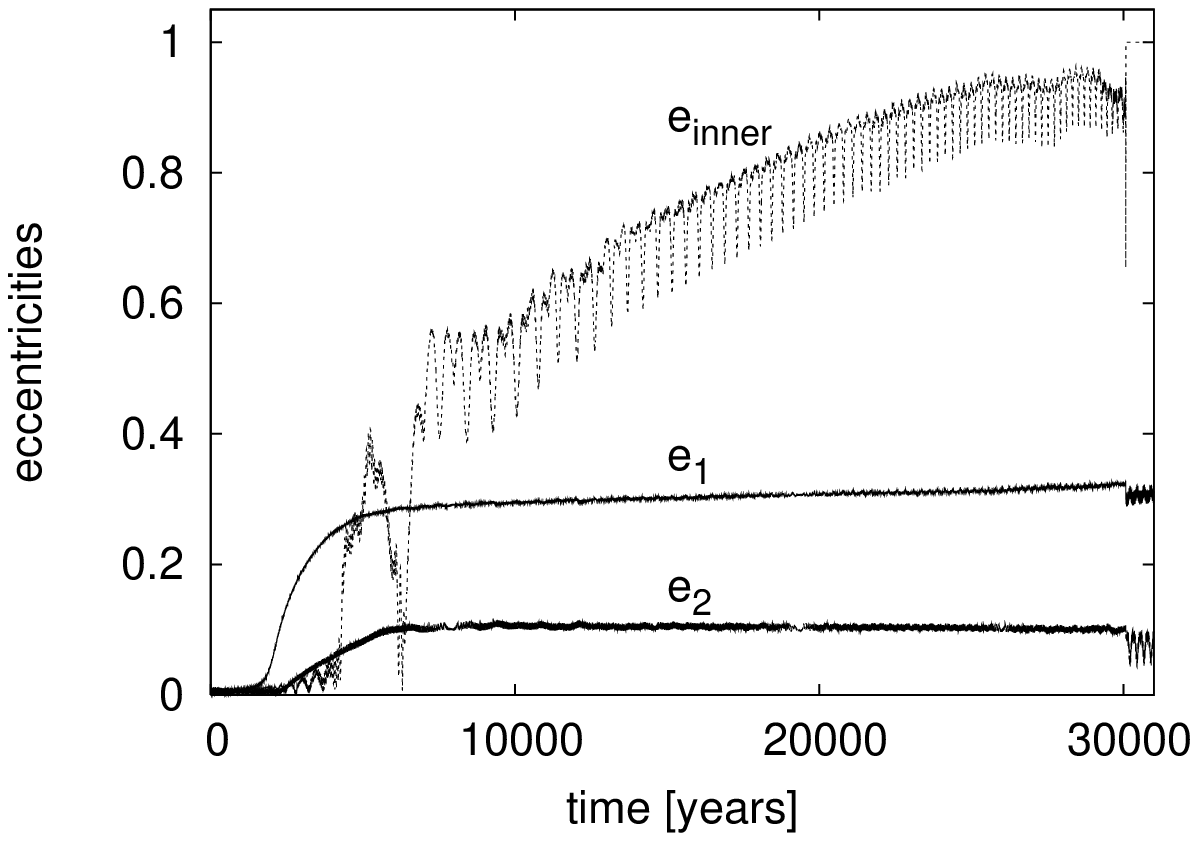} 
  \centering 
  \includegraphics[width=8cm]{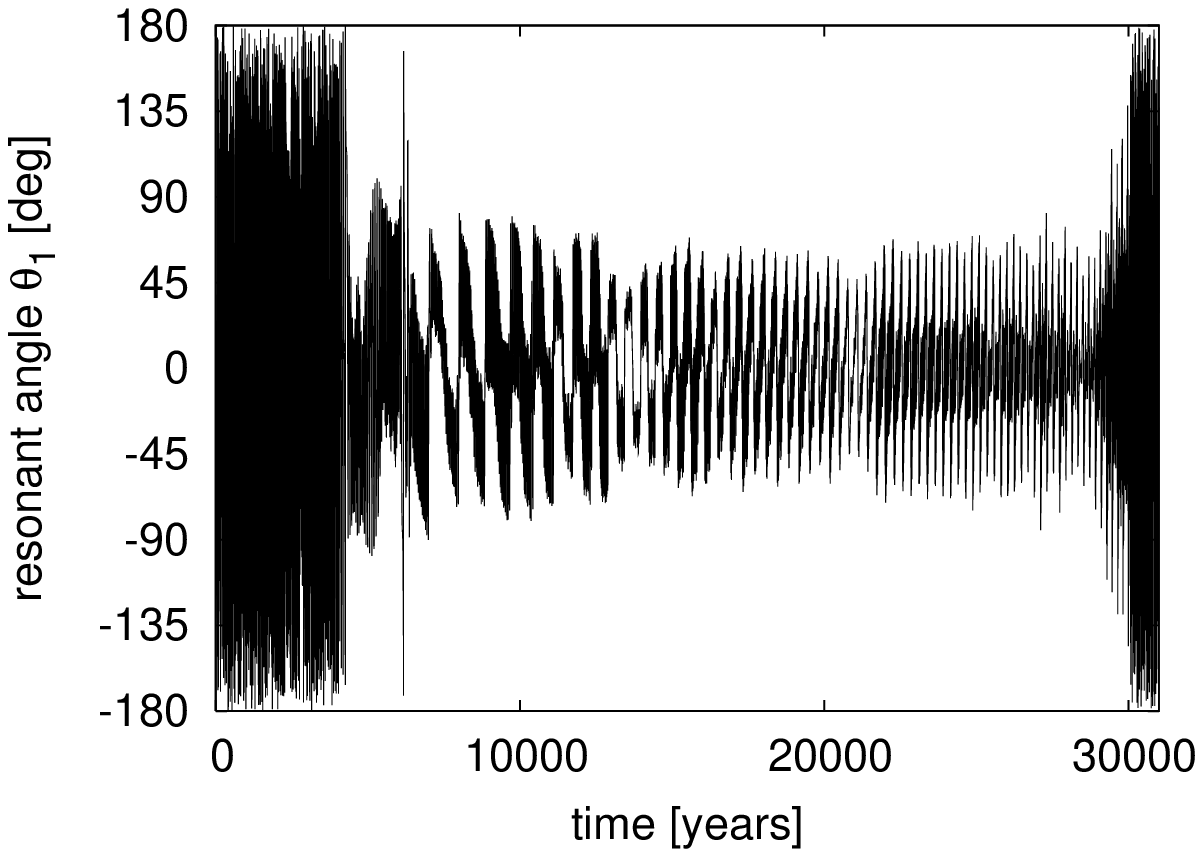} 
  \caption{{\bf Top}: The evolution of the eccentricities of the giant and the small mass 
  planets during a planetary migration when the small mass planet is also captured into a 
  $3:1$ resonance by the inner giant planet. In this case the resonant configuration is no 
  more protecting, the eccentricity of the small mass planets increases during the 
  migration and its orbit can cross the orbit of the inner giant planet. {\bf Bottom}: The 
  behavior of the resonant angles $\theta_1$ during the migration process. We note that 
  the other resonant angles (of the $3:1$ resonance) are circulating, thus the orbits of the 
  inner giant and the small mass inner planet are not in apsidal corotation.} 
         \label{fig:capt31noprot} 
\end{figure} 
 
\subsection{Scattering with a small mass inner planet} 
 
The behavior of the eccentricities of the giant planets around HD~73526 is very similar 
to that observed in the case of the system HD~128311 and $\upsilon$ Andromedae. In 
the case of $\upsilon$ Andromedae, \citet{Fordetal2005Nature} proposed that such a 
behavior is most likely the result of a planet-planet scattering event. Studying the 
system HD~128311, \citet{Sandor&Kley2006A&A} have demonstrated that the 
observed behavior of that system may be the consequence of a planet-planet scattering 
as well. In their recent paper \citet{Tinneyetal2006ApJ} have also suggested that the 
dynamical behavior of the system around HD~73526, might be the result of a dynamical 
scattering event. In what follows, we shall investigate whether the present behavior of 
HD~73526 can be modeled by such an effect.  
 
We have assumed that in addition to the giant planets, being already in a $2:1$ resonance and 
migrating towards the hosting star, a small mass planet ($\sim 10M_{\oplus}$) is also 
orbiting close to the hosting star in a quasi circular orbit. (This assumption could be 
realistic as the discovery of a $7.5M_{\oplus}$ planet around GJ~876 shows 
\citep{Riveraetal2005ApJ}) Initially, the giant planets are far enough from the small 
mass planet thus they do not influence significantly its motion. However, as the giant 
planets migrate inward approaching their present positions, they perturb the motion of 
the small planet. As our numerical experiments show at the corresponding ratio of the 
semi-major axes, the inner giant planet captures the small mass planet into a $3:1$ 
resonance. Once the capture into the $3:1$ resonance happens, two different scenarios 
have been detected: (i) the eccentricity of the small mass planet grows initially, then 
oscillates around the eccentricity of the inner giant planet, (ii) the eccentricity of the 
small mass planet grows during the whole migration process, which can result in asurvived 
possible close encounter between the inner small mass and the giant planet.  
 
In the case (i) the resonant angles of the $3:1$ resonance  
$\theta_1=3\lambda_1 - \lambda_{inn} - \varpi_1 - \varpi_{inn}$, 
$\theta_2=3\lambda_1 - \lambda_{inn} - 2\varpi_1$, and  
$\theta_3=3\lambda_1 - \lambda_{inn} -  2\varpi_{inn}$ librate around  
$\sim 40^{\circ}$, $\sim 150^{\circ}$, and $\sim 290^{\circ}$, respectively, while  
$\Delta \varpi^{*}=\varpi_1-\varpi_{inn}$ also librates around $\sim -110^{\circ}$ 
showing that the orbits of the inner giant and of the small mass planet are in an 
asymmetric apsidal corotation. (Here the index `1' refers to the inner giant planet, while 
the index `$inn$' to the small mass inner planet.) This resonant configuration protects 
the small mass inner planet from the close encounter (making impossible the 
planet-planet scattering), and stabilizes its orbit during the whole migration. The 
behavior of the eccentricities of the planets and the resonant angle $\theta_1$ (for the 
 $3:1$ resonance), $\Delta \varpi^{*}$ are shown in Figure \ref{fig:capt31prot}. 
 
In the case (ii) the inner giant planet also captures the small mass planet into a $3:1$ 
resonance, however, in this case only $\theta_1$ librates around $0^{\circ}$ (see the 
bottom panel of Figure \ref{fig:capt31noprot}), the other resonant angles are 
circulating, thus the orbits of the inner giant and the small mass planets are not in 
apsidal corotation. During the migration, the eccentricity of the small mass inner planet 
grows (shown in the top panel of Figure \ref{fig:capt31noprot}) resulting in a more 
and more elongated orbits. Thus eventually the orbit of the small mass planet will cross the orbit 
of the inner giant planet and between the planets a close encounter can happen. 
After the close encounter the small planet is either ejected from the system or pushed 
into a distant orbit from the hosting star. 

Having performed several numerical simulations, we have found that the outcome of the 
resonant capture described in the cases (i) and (ii), depends mainly on the initial positions 
of the inner giant and small mass planet, and it is not very sensitive on the mass of the
small inner planet (ranging between $0.03$ and $0.04 M_{Jup}$). It is very probably that 
prior to its secular increase, $e_{inn}$ goes through a sudden change in the family of 
corotation described and analysed recently also for the $3:1$ resonance by 
\citet{Michtchenkoetal2006CeMDA}. Thus a detailed analysis would certainly be useful to 
study the mechanism and the probability of the resonant capture into the $3:1$ resonance 
leading to either a protective behavior or a close encounter with the inner giant planet and a 
possible ejection of the small mass planet. However, at the present stage of our investigations 
we restrict ourselves only to demonstrate the mechanism that leads to the increase of the 
eccentricity of the small mass inner planet and to a planet-planet scattering event.  
 
In our numerical experiments we varied the mass of the small planet between $0.03$ 
and $0.04 M_{Jup}$  and the migration timescale as well. We have performed several 
simulations, and have found that once scattering happens, in the majority of the cases 
the system remains captured into the $2:1$ resonance. However, in some cases we 
have observed the complete destruction of the resonance, or even escape of the giant 
planets. We note that if the scattering takes place well before the disk dispersal 
process, the large oscillations in the eccentricities may be smoothed out due to the damping 
acting of the disk acting on the eccentricity of the outer giant 
planet. If the scattering happens during or after the disk dispersal, the oscillations of  
the eccentricities are preserved, and the system reaches its presently observed state.  
We have found that in order to observe the continuous increase of the small mass inner  
planet's eccentricity the migration of the giant planets should take sufficiently long time.   
 
Finally, the question may arise whether it is necessary to assume a kind of temporal  
synchronization between the termination of the migration and the scattering event. We have  
performed several numerical investigations and found in many cases that the inward migrating  
giant planets represent a perturbation strong enough to make the behaviour of the small mass 
inner planet unstable. However, the small planet survived for longer times, and the close  
encounter happened when the migration was already terminated. On the other hand there can  
also be a {\it natural} temporal synchronization. The migration of the giant planet should  
terminate when they reach their presently observed position. If a small mass planet orbits  
at a distance from the central star, which is needed for the resonant capture, its orbit  
will be captured into a $3:1$ resonance, and therefore its eccentricity pumped up making  
possible a close encounter with the inner giant planet, and finally, its ejection from the  
system. 
  
\begin{figure} 
  \centering 
  \includegraphics[width=8cm]{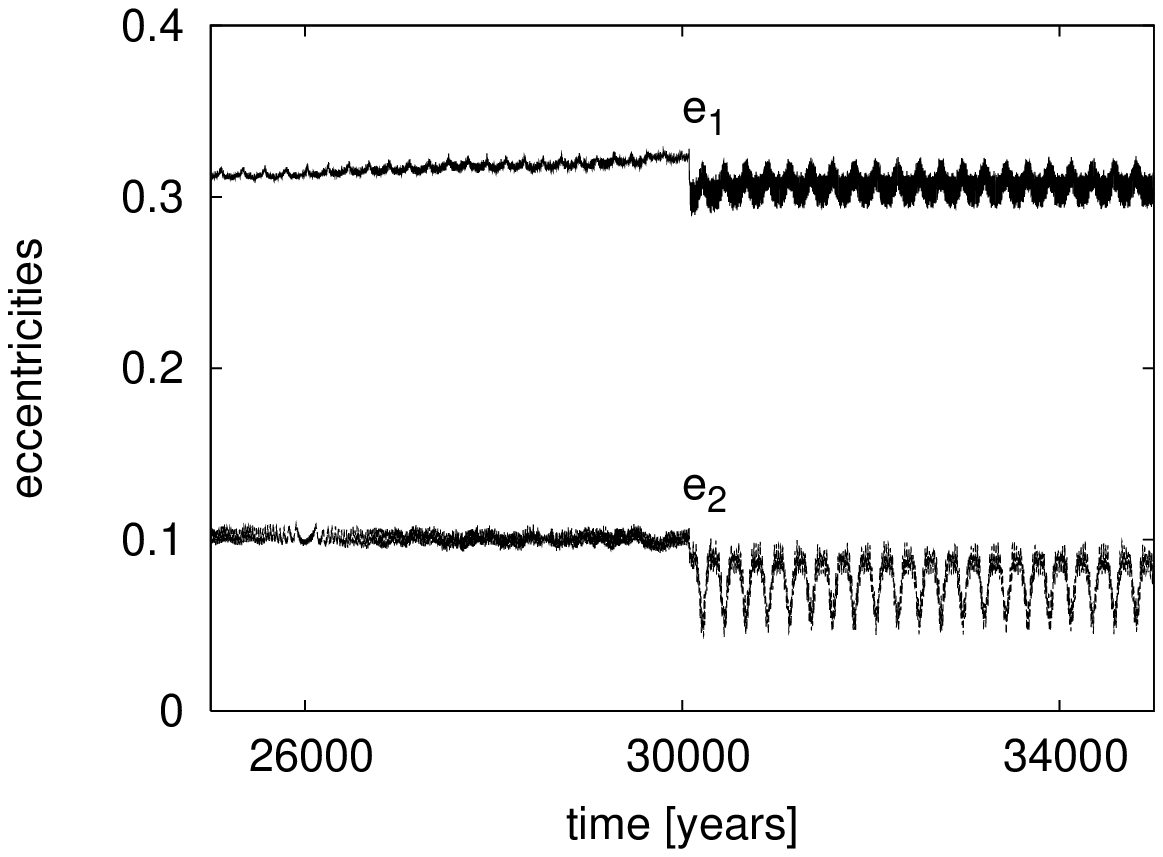} 
  \centering 
  \includegraphics[width=8cm]{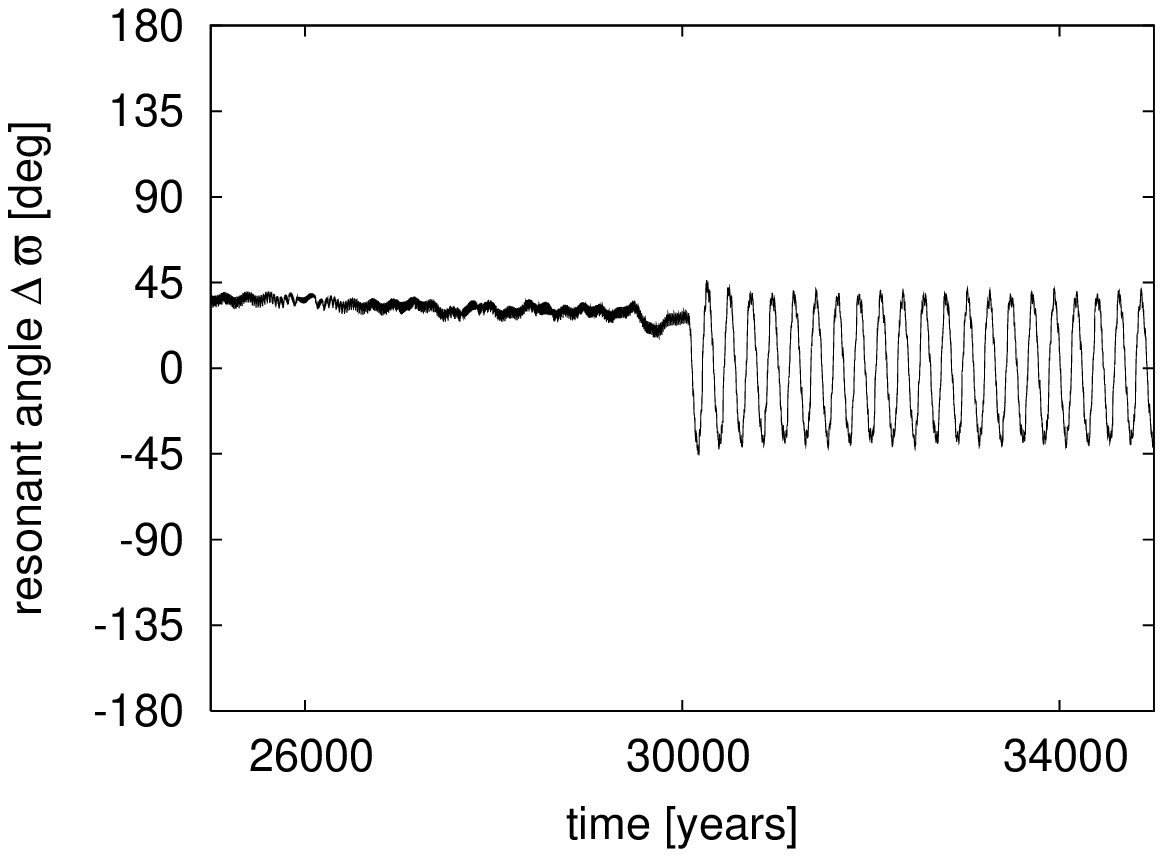} 
  \caption{{\bf Top}: The behavior of the eccentricities of the giant planets during  
  a planetary migration process before and after a planet-planet scattering event taking place 
  around $t_{sc}\approx3\times 10^4$ years. {\bf Bottom}: The behavior of 
  $\Delta\varpi$. The behavior of the eccentricities and $\Delta\varpi$ is very similar 
  of those of Fit 3.} 
         \label{fig:planetscattering} 
\end{figure} 
 
In Figure \ref{fig:planetscattering}, which is the magnification of Figure 
\ref{fig:capt31noprot}, we show the typical behavior of the eccentricities of the giant 
planets, and $\Delta\varpi$ during a migration process before and after a planet-planet 
scattering. The giant planets are started initially on distant orbits having initial 
semi-major axes $a_1=2$ AU and $a_2=4$ AU. The $e$-folding time of the outer 
giant's semi-major axis is $\tau_{a_{2}}= 10^4$ years and $K=15$.  
 
The mass of the inner 
small planet is $M_{inn}=0.027M_{Jup}$, and it orbits at $a_{inn}=0.9$ AU from 
the hosting star in a nearly circular orbit. The scattering occurs around $t_{sc}=3\times 
10^4$ years. We applied also a linear dispersal process of the protoplanetary disk 
taking place between $2.9\times 10^4$ and $3.1\times 10^4$ years. Thus the 
oscillations in the eccentricities induced by the scattering event are not damped out 
anymore. Since the giant planets are still in a (protective) 2:1 resonance their stability is 
preserved for the whole life-time of the system. We note that in contrary to the case 
HD~128311, during modeling the formation of HD~73526 we have not observed the 
breaking of the apsidal corotation. It is highly probable that due to the large planetary 
masses and relatively small semi-major axes, the system is more fragile than 
HD~128311, and a sufficiently strong perturbation which could lead to the breaking 
of the apsidal corotation may destroy the whole system. However, after the scattering 
event the resonant angle $\Delta\varpi$ shows large oscillations around $0^{\circ}$. 
 
\begin{figure} 
  \centering 
  \includegraphics[width=8cm]{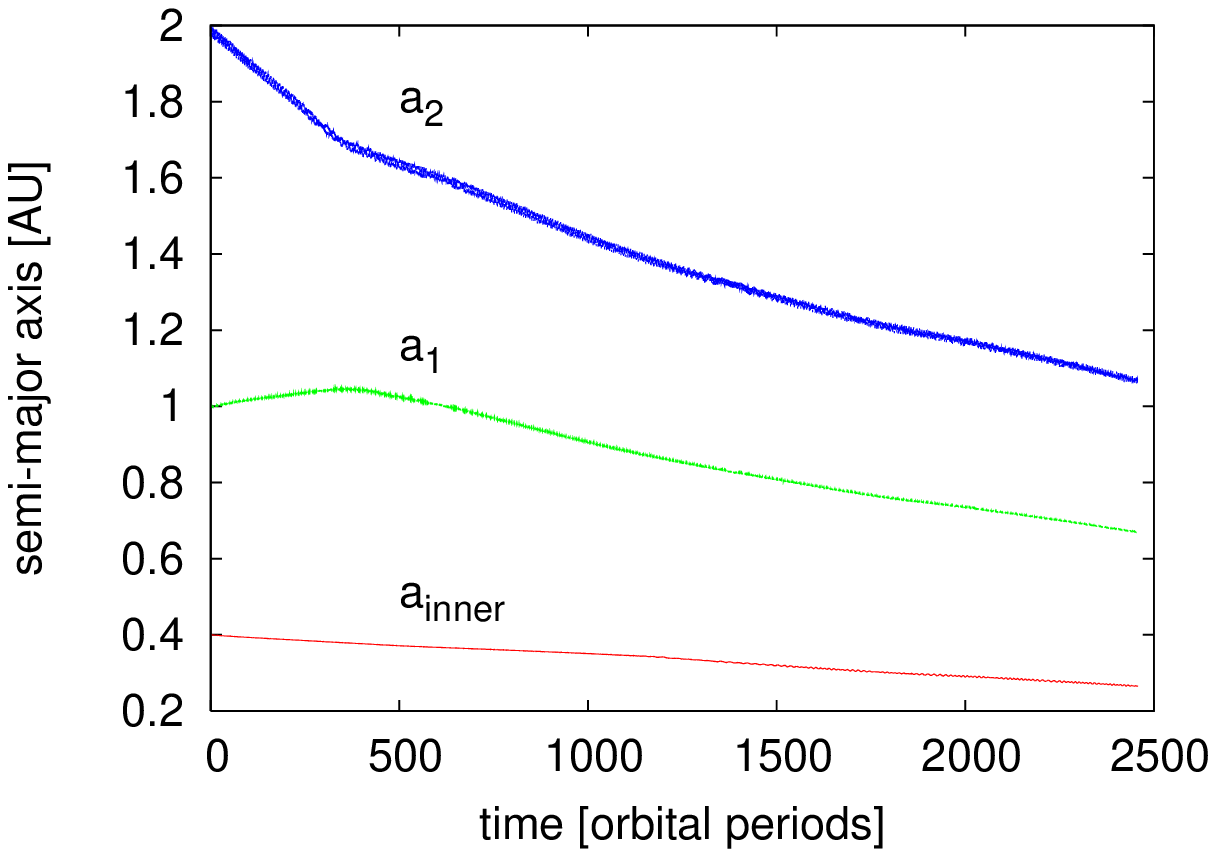} 
  \centering 
  \includegraphics[width=8cm]{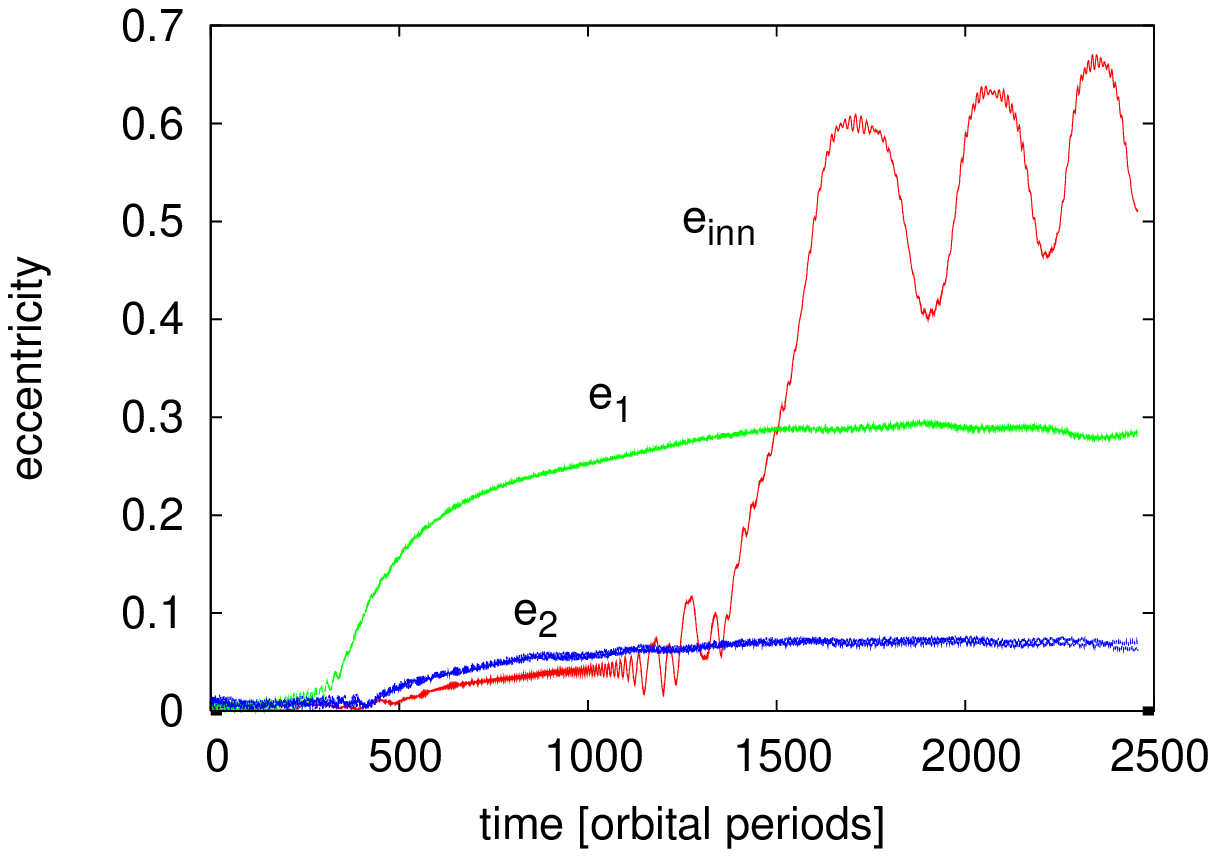}  
  \caption{ 
  Full hydrodynamical evolution of two embedded massive planets and an 
   small additional small mass inner planet. 
  {\bf Top}: The behavior of the semimajor axis of the three planets. 
  {\bf Bottom}: The behavior of the eccentricities. 
  } 
         \label{fig:threeplanet} 
\end{figure} 
 
We note that in the above simulations (gravitational three-body problem with  
dissipative forces) the small mass inner planet has not been affected by the inner disk.  
In this case we used a model without an inner disk, which may also be realistic in some cases. 
However the damping ratio $K=15$ is a bit high; smaller $K$ would result in higher  
eccentricity of the inner giant planet. As we have already demonstrated in 4.1,  
the presence of an inner disk can damp effectively the eccentricity of the inner giant planet.   
In order to investigate the effect of the inner disk on the small mass planet we have also  
performed full hydrodynamical simulations using the same initial parameters as described in 4.1,  
but in addition to the giant planets, a $M_{inn}=10M_{\oplus}$ inner planet has also been embedded  
in the inner disk and started from $0.4$AU. According to the theoretical expectations, its  
migration rate (being in linear regime with no gap opening) is smaller than that of a more  
massive Jupiter-like planet. Thus the small mass planet has been captured into a $3:1$  
resonance resulting in an increase of its eccentricity.  
In Fig.~\ref{fig:threeplanet} we display the time evolution of the orbital elements 
of the three planets. At $t=400$ orbital periods (of the inner giant planet) the two massive 
planets capture each other in a $2:1$ resonance slowing down the inward migration which 
nevertheless remains faster than that of the inner planet which is consequently captured at a 
later time. We should also note that due to the very long computational time, we have not 
followed  the complete evolutionary way of the small mass inner planet. On the other hand it 
is very probable, that the increase of its eccentricity will result in a scattering with the 
inner giant planet, and finally, ejection from the system.  
Similarly to the eccentricity damping mechanism, the final behavior of the inner small  
mass planet will also be the subject of a more detailed analysis investigating inner disks  
with different physical properties. However, our results to date clearly show that even in  
the presence of an inner disk the eccentricity excitation mechanism of the small mass inner  
planet (through resonant capture) works well, and enables scattering between the small mass 
and the inner giant planet.  
 
\section{Conclusion} 
 
In this paper we have investigated different evolutionary scenarios, which may have  
led to the observed behavior of the resonant system HD~73526. According to our 
numerical integration, the original orbit fit provided by \citet{Tinneyetal2006ApJ} 
results in weakly chaotic motion. Thus, we have derived four sets of orbital elements, 
which fit very well to the observed radial velocity measurements, and also provide, 
through a protective 2:1 resonance, stable motion for the giant planets during the whole 
life-time of the system. Two of these fits yield very large variations of the eccentricities 
of the giant planets, in one case the giant planets are not even in apsidal resonance. In 
the case of the other two fits the variations of the eccentricities are more moderate, and 
beside the mean motion resonance, the giant planets are in apsidal corotation.  
 
Since the behavior of the giant planet's eccentricities in the system HD~128311 is very 
similar to those in HD~73526, we have investigated whether this system may also be 
the result of a mixed formation scenario incorporating a slow migration followed by a 
perturbative event. Similarly to the formation of the system HD~128311, we have 
assumed that the two giant planets have been formed far from their hosting star, and 
migrated inward due to a planet-disk gravitational interaction. During this migration the 
planets have been locked in a protective 2:1 resonance.  
 
Performing full hydrodynamical simulations of two embedded giant planets in a 
protoplanetary disk we have found that the presence of an inner disk can 
damp efficiently the eccentricity of the inner giant planet resulting in eccentricity values 
being in good agreement with the observations. This mechanism has not been 
investigated yet, and may offer a real possibility to avoid the use of very high values of 
the ratio $K$, for example in modeling the resonant system around GJ~876.  
We have additionally found that the effect of an inner disk can be modeled very 
conveniently by using non-conservative forces. We have investigated three convergent 
migration scenarios that can lead the system to the deep $2:1$ 
resonance, which are all physically realistic. 
 
Similarly to HD~128311 we have assumed that the system HD~73526 suffered a 
sudden perturbation near the end of the migration of the giant planets. This perturbation 
could either be the quick termination of the migration, which may possibly be induced 
by an inner rim of the disk and an empty region inside of it, as indicated by some 
observations of young stars, or a planet-planet scattering event. We have analyzed the 
results of an encounter of one of the giant planets with a relatively small $\sim 10 
M_{\oplus}$ planet, which orbits around the central star at a closer distance ($a<1$ 
AU), and which is ejected from the system, or thrown to a very distant orbit after the 
scattering event. 
 
Investigating the effect of the disk's sudden dispersal, we have found 
that only a very short disk 
dispersal time $\Delta t_{\mathrm disp}=50$ years will induce the observed 
oscillations in the eccentricities and in the resonant angles. The reason of this behavior 
is that during the migration of the giant planets the system tends to an asymmetric 
apsidal corotation with $\Delta\varpi \neq 0^{\circ}$.  At moment of the sudden termination of 
the migration $\Delta\varpi\approx  50^{\circ}$, and $\Delta\varpi$ 
will continue to librate with 
this amplitude around its equlibrium ($0^{\circ}$). If $\Theta_1$ is also different from 
$0^{\circ}$, it will also librate around $0^{\circ}$ after the scattering event. We note 
however that in order to obtain the present behavior of the system we need relatively 
fast migration of the giant planets $\tau_{a_2}=4\times 10^3$ years and $K=15$. 
 
Studying the planet-planet scattering event we have found that the inner giant planet 
always captures the small mass inner planet in a $3:1$ resonance. Then the further 
evolution of the system depends on whether the orbits of the giant and the small mass 
planet are in apsidal corotation after the resonant capture. The apsidal corotation protects 
the orbit of the small mass planet from the close encounter with the inner giant planet, 
thus inhibits a planet-planet scattering event. The lack of the apsidal corotation results 
in a continuous increase of the small mass planet's eccentricity. Thus its orbit crosses 
the orbit of the giant planet and a close encounter can take place 
between them. Afterwards, if the system survives that perturbative event, the eccentricities of 
the giant planets behave qualitatively similar to those of our Fit~3, 
in agreement with the observed radial velocity observations of HD~73536.  
 
We can conclude that based on radial velocity measurements to date, it is probable that 
during the formation and early evolution of the systems HD~73526 and HD~128311 
the adiabatic migration, which has brought the systems into a protective $2:1$ resonance, 
has been followed by a sudden perturbative event. This perturbative event could be 
either the fast dispersal of the circumstellar disk or a planet-planet scattering. Thus the 
dynamics of the system HD~73526 confirm that planetary migration followed by 
sudden perturbations (e.g. planet-planet scattering) play an important role in shaping the 
formation and early dynamics of resonant planetary systems.   
 
\begin{acknowledgements} 
      We thank Prof. C. Beaug\'e for his careful reading of the manuscript and his useful
      comments and suggestions. This work has been supported by the Hungarian Scientific 
      Research Fund (OTKA) under the grant D048424, and by the German Research 
      Foundation (DFG) under grant 436 UNG 17/1/07.   
\end{acknowledgements} 
 
\bibliographystyle{aa} 

\begin{thebibliography}{23}
\expandafter\ifx\csname natexlab\endcsname\relax\def\natexlab#1{#1}\fi

\bibitem[{{Beaug{\'e}} {et~al.}(2006){Beaug{\'e}}, {Michtchenko}, \&
  {Ferraz-Mello}}]{Beaugeetal2006MNRAS}
{Beaug{\'e}}, C., {Michtchenko}, T.~A., \& {Ferraz-Mello}, S. 2006, \mnras,
  365, 1160

\bibitem[{{Bois} {et~al.}(2003){Bois}, {Kiseleva-Eggleton}, {Rambaux}, \&
  {Pilat-Lohinger}}]{Boisetal2003ApJ}
{Bois}, E., {Kiseleva-Eggleton}, L., {Rambaux}, N., \& {Pilat-Lohinger}, E.
  2003, \apj, 598, 1312

\bibitem[{{Calvet} {et~al.}(2005){Calvet}, {D'Alessio}, {Watson},
  {Franco-Hern{\'a}ndez}, {Furlan}, {Green}, {Sutter}, {Forrest}, {Hartmann},
  {Uchida}, {Keller}, {Sargent}, {Najita}, {Herter}, {Barry}, \&
  {Hall}}]{Calvetetal2005ApJ}
{Calvet}, N., {D'Alessio}, P., {Watson}, D.~M., {et~al.} 2005, \apjl, 630, L185

\bibitem[{{Crida} {et~al.}(2007){Crida}, {Morbidelli}, \&
  {Masset}}]{2007A&A...461.1173C}
{Crida}, A., {Morbidelli}, A., \& {Masset}, F. 2007, \aap, 461, 1173

\bibitem[{{D'Alessio} {et~al.}(2005){D'Alessio}, {Hartmann}, {Calvet},
  {Franco-Hern{\'a}ndez}, {Forrest}, {Sargent}, {Furlan}, {Uchida}, {Green},
  {Watson}, {Chen}, {Kemper}, {Sloan}, \& {Najita}}]{DAlessioetal2005ApJ}
{D'Alessio}, P., {Hartmann}, L., {Calvet}, N., {et~al.} 2005, \apj, 621, 461

\bibitem[{{de Val-Borro} {et~al.}(2006){de Val-Borro}, {Edgar}, {Artymowicz},
  {Ciecielag}, {Cresswell}, {D'Angelo}, {Delgado-Donate}, {Dirksen}, {Fromang},
  {Gawryszczak}, {Klahr}, {Kley}, {Lyra}, {Masset}, {Mellema}, {Nelson},
  {Paardekooper}, {Peplinski}, {Pierens}, {Plewa}, {Rice}, {Sch{\"a}fer}, \&
  {Speith}}]{2006MNRAS.370..529D}
{de Val-Borro}, M., {Edgar}, R.~G., {Artymowicz}, P., {et~al.} 2006, \mnras,
  370, 529

\bibitem[{{Ford} {et~al.}(2005){Ford}, {Lystad}, \&
  {Rasio}}]{Fordetal2005Nature}
{Ford}, E.~B., {Lystad}, V., \& {Rasio}, F.~A. 2005, \nat, 434, 873

\bibitem[{{Hayashi}(1981)}]{Hayashi1981PThPS}
{Hayashi}, C. 1981, Progress of Theoretical Physics Supplement, 70, 35

\bibitem[{{Kley}(1999)}]{Kley1999MNRAS}
{Kley}, W. 1999, \mnras, 303, 696

\bibitem[{{Kley}(2000)}]{Kley2000MNRAS}
{Kley}, W. 2000, \mnras, 313, L47

\bibitem[{{Kley} {et~al.}(2005){Kley}, {Lee}, {Murray}, \&
  {Peale}}]{Kleyetal2005A&A}
{Kley}, W., {Lee}, M.~H., {Murray}, N., \& {Peale}, S.~J. 2005, \aap, 437, 727

\bibitem[{{Kley} {et~al.}(2004){Kley}, {Peitz}, \& {Bryden}}]{Kleyetal2004A&A}
{Kley}, W., {Peitz}, J., \& {Bryden}, G. 2004, \aap, 414, 735

\bibitem[{{Lee}(2004)}]{Lee2004ApJ}
{Lee}, M.~H. 2004, \apj, 611, 517

\bibitem[{{Lee} \& {Peale}(2002)}]{Lee&Peale2002ApJ}
{Lee}, M.~H. \& {Peale}, S.~J. 2002, \apj, 567, 596

\bibitem[{{Masset} {et~al.}(2006){Masset}, {Morbidelli}, {Crida}, \&
  {Ferreira}}]{2006ApJ...642..478M}
{Masset}, F.~S., {Morbidelli}, A., {Crida}, A., \& {Ferreira}, J. 2006, \apj,
  642, 478
  
\bibitem[{{Michtchenko} {et~al.}(2006){Michtchenko}, {Beaug{\'e}}, \&
  {Ferraz-Mello}}]{Michtchenkoetal2006CeMDA}
{Michtchenko}, T.~A., {Beaug{\'e}}, C., \& {Ferraz-Mello}, S. 2006, Celestial
  Mechanics and Dynamical Astronomy, 94, 411
  
\bibitem[{{Perryman}(2000)}]{Perryman2000RPPh}
{Perryman}, M.~A.~C. 2000, Reports of Progress in Physics, 63, 1209

\bibitem[{{Pollack}(1984)}]{Pollack1984ARA&A}
{Pollack}, J.~B. 1984, \araa, 22, 389

\bibitem[{{Psychoyos} \&
  {Hadjidemetriou}(2005)}]{Psychoyos&Hadjidemetriou2005CeMDA}
{Psychoyos}, D. \& {Hadjidemetriou}, J.~D. 2005, Celestial Mechanics and
  Dynamical Astronomy, 92, 135

\bibitem[{{Rivera} {et~al.}(2005){Rivera}, {Lissauer}, {Butler}, {Marcy},
  {Vogt}, {Fischer}, {Brown}, {Laughlin}, \& {Henry}}]{Riveraetal2005ApJ}
{Rivera}, E.~J., {Lissauer}, J.~J., {Butler}, R.~P., {et~al.} 2005, \apj, 634,
  625

\bibitem[{{S{\'a}ndor} {et~al.}(2004){S{\'a}ndor}, {{\'E}rdi}, {Sz{\'e}ll}, \&
  {Funk}}]{Sandoretal2004CeMDA}
{S{\'a}ndor}, Z., {{\'E}rdi}, B., {Sz{\'e}ll}, A., \& {Funk}, B. 2004,
  Celestial Mechanics and Dynamical Astronomy, 90, 127

\bibitem[{{S{\'a}ndor} \& {Kley}(2006)}]{Sandor&Kley2006A&A}
{S{\'a}ndor}, Z. \& {Kley}, W. 2006, \aap, 451, L31

\bibitem[{{Tinney} {et~al.}(2006){Tinney}, {Butler}, {Marcy}, {Jones},
  {Laughlin}, {Carter}, {Bailey}, \& {O'Toole}}]{Tinneyetal2006ApJ}
{Tinney}, C.~G., {Butler}, R.~P., {Marcy}, G.~W., {et~al.} 2006, \apj, 647, 594

\bibitem[{{Vogt} {et~al.}(2005){Vogt}, {Butler}, {Marcy}, {Fischer}, {Henry},
  {Laughlin}, {Wright}, \& {Johnson}}]{Vogtetal2005ApJ}
{Vogt}, S.~S., {Butler}, R.~P., {Marcy}, G.~W., {et~al.} 2005, \apj, 632, 638

\end{thebibliography}
\end{document}